\documentclass[pra,a4paper,showpacs,twocolumn,superscriptaddress,longbibliography]{revtex4-1}

\usepackage{amssymb}
\usepackage{amsmath}
\usepackage{amsfonts}
\usepackage{graphicx}
\usepackage{bm}
\usepackage{color}
\usepackage{multirow}
\usepackage{natbib}
\usepackage{hyperref}

\DeclareMathOperator{\Tr}{Tr}
\DeclareMathOperator{\spp}{sp}
\DeclareMathOperator{\sgn}{sgn}
\DeclareMathOperator{\liq}{li_2}

\DeclareMathOperator{\im}{Im}
\DeclareMathOperator{\re}{Re}

\begin{document}

\title{Mesoscopic fluctuations of the single-particle Green's function at Anderson transitions with Coulomb interaction}

\author{E. V. Repin}

\affiliation{Delft University of Technology, The Netherlands}

\affiliation{L.D. Landau Institute for Theoretical Physics, Kosygina
  street 2, 117940 Moscow, Russia}

\author{I. S.~Burmistrov}

\affiliation{L.D. Landau Institute for Theoretical Physics, Kosygina
  street 2, 117940 Moscow, Russia}
  
\affiliation{Moscow Institute of Physics and Technology, 141700 Moscow, Russia}


\begin{abstract}
Using the two-loop analysis and the background field method we demonstrate that the local pure scaling operators without derivatives in the Finkel'stein nonlinear sigma model can be constructed by straightforward generalization of the corresponding operators for the noninteracting case. These pure scaling operators demonstrate multifractal behavior and describe mesoscopic fluctuations of the single-particle Green's function. We determine anomalous dimensions of all such pure scaling operators in the interacting theory within the two-loop approximation.  
\end{abstract}

\pacs{
72.15.Rn , \,
71.30.+h , \,
73.43.Nq 	\,
}

\maketitle

\section{Introduction}

Anderson transition is  a quantum phase transition (QPT) driven by disorder.  During more than half a century of a research after the seminal paper \cite{Anderson58}, a vast knowledge on Anderson transitions has been accumulated (see Ref. \cite{EversMirlin} for a review). Typically, this transition separates a metallic phase and an Anderson insulator. In the presence of nontrivial topology, e.g. as in the case of integer quantum Hall effect or time-reversal topological insulators, Anderson transition occurs between distinct topological phases. Similar to an ordinary QPT, Anderson transition is characterized by a divergent correlation/localization length $\xi$ and by critical scaling of physical observables. A striking feature of Anderson transition is existence of strong mesoscopic fluctuations of electron wave functions which lead to multifractal behavior at criticallity \cite{Wegner1980,Castellani1986,Lerner1988}. The consequence of multifractality of wave functions is the following scaling behavior of moments of the local density of states (LDOS) with a system size $L$:
\begin{equation}
\frac{\langle \rho^q(E,\bm{r})\rangle_{\rm dis}}{\langle \rho(E,\bm{r})\rangle^q_{\rm dis}}  \propto L^{-\Delta_q} ,
\label{eq:RR1}
\end{equation}
where the critical exponent $\Delta_q\leqslant 0$ is a nonlinear function of nonnegative integer $q$. 

For a long time multifractality in disordered systems remains a theoretical concept which was studied in 
numerical experiments only (see Refs. \cite{Janssen,Huckestein,EversMirlin} for a review). Recently, situation was changed dramatically. On the one hand, it was realized that multifractality of LDOS leads to strong enhancement of superconducting transition temperature \cite{FeigelmanYuzbashyan2007,FeigelmanCuevas2010,BurmistrovGornyiMirlin2012,Burmistrov2015b,DellAnna}, responsible for instabilities of surface states in topological superconductors \cite{Foster2012,Foster2014}, results in strong mesoscopic fluctuations of Kondo temperature \cite{Kettemann2006,Micklitz2006,Kettemann2007}, and affects the Anderson orthogonality catastrophe \cite{Kettemann2016}. On the other hand, a signature of multifractality has been found experimentally in an electron system in diluted magnetic semiconductor Ga$_{1-x}$Mn$_x$As \cite{Richardella}, in ultrasound waves propagating through a system of randomly packed Al beads \cite{Faez2009}, in light waves spreading in an array of dielectic nanoneedles \cite{Mascheck2012}. 

Typically, relations of a type of Eq. \eqref{eq:RR1} suffer from existence of subleading corrections. The moments of  the LDOS are  remarkable due to absence of such corrections to scaling in Eq. \eqref{eq:RR1}. In fact, there are many more correlation functions which at Anderson transition demonstrate pure scaling behavior similar to Eq. \eqref{eq:RR1} with negative critical exponents \cite{Wegner1986}. Recently, a recipe for construction of  such pure scaling observables from disorder-averaged combinations of electron wave functions (or, alternatively, single-particle Green's functions) at different spatial points was proposed \cite{Gruzberg}. 

The theoretical framework for description of Anderson transition is provided by the nonlinear sigma model (NLSM) \cite{Wegner1979}. Within NLSM approach the critical exponents $\Delta_q$ are determined by the anomalous dimensions of certain pure scaling operators without spatial derivatives. The set of all such pure scaling operators (with negative and positive critical exponents) has been found with the help of the group-theoretical treatment of the NLSM manifold \cite{Wegner1986}. Remarkably, in Ref. \cite{Gruzberg} the exact symmetry relations between critical exponents of these pure scaling operators have been proven. 

The above progress in understanding of multifractal behavior of electron wave functions have been developed for  Anderson transitions in the absence of interactions. Metal-insulator transitions can occur  in the presence of both disorder and electron-electron interaction \cite{McMillan,Finkelstein1983a,Castellani1984}. In this case they are usually termed as Mott-Anderson transitions (see Refs. \cite{Finkelstein1990,BelitzKirkpatrick1994} for a review).  Typically, not only Coulomb but even short-range electron-electron interaction is a relevant perturbation in the renormalization-group (RG) sense for a noninteracting fixed point describing Anderson transition. Then Mott-Anderson transition corresponds to an  interacting fixed point 
for which, in general, a set of critical exponents is different from the set for a noninteracting case. 

Until recently, a fate of multifractality at Mott-Anderson transitions has been not known. In Refs. \cite{Burmistrov2013,Burmistrov2014,Burmistrov2015a}, it was demonstrated within NLSM treatment in $d=2+\epsilon$ dimensions that the scaling of LDOS, Eq. \eqref{eq:RR1}, exists at metal-insulator transitions in the presence of Coulomb interaction. In agreement with general expectations, critical exponents $\Delta_q$ differ from their values for a noninteracting fixed point. We emphasize that survival of multifractality in LDOS in the case of Coulomb interaction is not obvious a priori because of the so-called zero-bias anomaly, i.e. strong suppression of the disorder-averaged LDOS at the Fermi energy \cite{Altshuler1979b,*Altshuler1980,AAbook,Efros1975,SEbook}. The statement of Refs. \cite{Burmistrov2013,Burmistrov2014,Burmistrov2015a} about existence of multifractality of LDOS in the presence of Coulomb interaction is in agreement with an earlier numerical analysis in the framework of functional density theory \cite{Slevin2012,*Slevin2014}, and by the Hartree-Fock simulation of the problem \cite{Amini2014}. The consideration of Refs. \cite{Burmistrov2013,Burmistrov2014,Burmistrov2015a} were limited to moments of the LDOS which, as it was shown there, correspond to pure scaling operators of the Finkel'stein NLSM.  

In this paper, we answer a more general question: \emph{what are pure scaling operators of the Finkel'stein NLSM which describe mesoscopic fluctuations of the single-particle Green's function at Mott-Anderson transition?} Quite surprisingly, we find that these pure scaling operators of the Finkel'stein NLSM can be constructed by straightforward generalization of pure scaling operators without derivatives known for the noninteracting NLSM. Within the two-loop approximation we determine anomalous dimensions of all such pure scaling operators in the interacting theory. We demonstrate that anomalous dimensions are modified by the presence of interaction.   

The paper is organized as follows. In Sec. \ref{Sec:Form} we remind formalism of the Finkel'stein NLSM. 
The procedure of construction of local operators without spatial derivatives will be explained in Sec. \ref{Sec:Oper}. In Sec. \ref{Sec:2loop} we present the results of two-loop renormalization of local operators without spatial derivatives and demonstrate that the pure scaling operators in the Finkel'stein NLSM
 are straightforward generalization of the pure scaling operators of the noninteracting theory. The two-loop arguments are supported by the background field renormalization method in Sec. \ref{Sec:BFM}. We conclude the paper with the discussion of experimental relevance of our results and summary of our findings (Sec. \ref{Sec:Sum}). Some necessary details of two-loop analysis and background field renormalization are given in Appendix.

\section{Nonlinear sigma model formalism\label{Sec:Form}}

The effective field theory in the case of preserved spin-rotational and time-reversal symmetries is defined in a standard way \cite{Finkelstein1990,BelitzKirkpatrick1994}. It is formulated for a matrix field $Q$ which takes values in a symmetric space $G/K$ with $G={\rm Sp}(2N)$ and $K={\rm Sp}(N)\times {\rm Sp}(N)$. The rank of the symplectic group is given by $N= 4 N_r N_m$ where $N_r$ denotes a number of replica  and $N_m$ stands for a number of Matsubara frequencies involved. The factor $4$ corresponds to the spin and Nambu (particle-hole) spaces. The effective action can be written as follows
\begin{align}
S=& -\frac{g}{32} \int d\bm{r} \Tr (\nabla Q)^2 + 4\pi T Z_\omega \int d\bm{r} \Tr \eta Q \notag \\
& - \frac{\pi T}{4} \sum_{\alpha,n,r,j}\Gamma_j 
\int d\bm{r} \Tr \Bigl [I_n^\alpha t_{rj} Q\Bigr ]  \Tr \Bigl [I_{-n}^\alpha t_{rj} Q\Bigr ] .
\label{eq:NLSM}
\end{align}
Here we introduce the following matrices
\begin{equation}
\eta_{nm}^{\alpha\beta}=n \, \delta_{nm}\delta^{\alpha\beta} t_{00},  \quad
(I_k^\gamma)_{nm}^{\alpha\beta}=\delta_{n-m,k}\delta^{\alpha\beta}\delta^{\alpha\gamma} t_{00} ,
\end{equation}
where $\alpha,\beta = 1,\dots, N_r$ stands for replica indices and indices $n,m$ correspond to the
Matsubara fermionic frequencies $\varepsilon_n = \pi T (2n+1)$ ($T$ stands for the temperature.). The sixteen matrices $t_{rj}$ act 
in a tensor product of the spin (subscript $j$) and Nambu  (subscript $r$) spaces. The indices $r$ and $j$ of matrices $t_{rj}$ indicate its decomposition by the basis of tensor products of the unit and Pauli matrices in these spaces, i.e. 
\begin{equation}
\label{trj}
t_{rj} = \tau_r\otimes s_j, \qquad r,j = 0,1,2,3  .
\end{equation}
Here matrices $\tau_0$ and $s_0$ stand for the $2\times 2$ unit matrices and
\begin{equation}
\tau_1/s_1  = \begin{pmatrix}
0 & 1\\
1 & 0
\end{pmatrix}, \, \tau_2/s_2 = \begin{pmatrix}
0 & -i\\
i & 0
\end{pmatrix}, \, \tau_3/s_3 = \begin{pmatrix}
1 & 0\\
0 & -1
\end{pmatrix} . \notag
\end{equation}
The total (including spin) dimensional  (in units $e^2/h$) Drude conductivity is denoted by $g$. The interaction amplitudes $\Gamma_j$ (for the singlet channel, $\Gamma_0 = \Gamma_s$, and for the triplet channel, $\Gamma_1=\Gamma_2=\Gamma_3=\Gamma_t$) describe electron-electron interaction in the particle-hole channel. In this paper we neglect the interaction in the Cooper channel. The parameter $Z_\omega$ takes into account nontrivial frequency renormalization under the RG flow \cite{Finkelstein1983a}. The bare value of $Z_\omega$ is unity. For computation of physical observables with the  NLSM action \eqref{eq:NLSM} one needs to take two limits: $N_m \to \infty$ and $N_r\to 0$. We note that the former limit  
is tricky and should be performed in a way consistent with the gauge invariance (see Ref. \cite{Baranov1999a} for details).

Matrix $Q(\bm{r})$ describes local rotations around the spatially independent matrix $\Lambda$:
\begin{equation}
Q=\mathcal{T}^{-1} \Lambda \mathcal{T}, \quad \Lambda_{nm}^{\alpha\beta} = \sgn \varepsilon_n \, \delta_{nm} \delta^{\alpha\beta}t_{00},
\end{equation}
where the matrices $\mathcal{T} \in G$ obey the following relations:
\begin{equation}
C (\mathcal{T}^{-1})^{\rm T} = \mathcal{T} C,\qquad \mathcal{T}^{\rm T} C = C \mathcal{T}^{-1} . \label{TC}
 \end{equation}
Here $C = i t_{12}$ and $\mathcal{T}^{\rm T}$ denotes the matrix transpose of $\mathcal{T}$.  Therefore the matrix $Q$ is subjected to the local nonlinear constraint, $Q^2(\bm{r})=1$, satisfies $\Tr Q=0$ and charge-conjugate condition,
\begin{equation}
 Q=Q^\dag = C^T Q^T C .
 \label{TC2}
\end{equation}

\section{Local operators without derivatives\label{Sec:Oper}}

The simplest local operator without derivatives is given as
\begin{equation}
\mathcal{K}_1(E) = \frac{1}{4} \re \mathcal{P}_1^+(E) ,
\label{eq:K1}
\end{equation}
where the retarded correlation function $\mathcal{P}_1^+(E)$ can be obtained from its Matsubara counterpart
\begin{equation}
\mathcal{P}_1(i\varepsilon_n) = \spp \langle Q_{nn}^{\alpha\alpha}\rangle 
\label{eq:P1M}
\end{equation}
after standard analytic continuation, $i\varepsilon_n \to E+i0^+$. Here the  symbol $\spp$ denotes the trace over spin and Nambu spaces. There is no summation over a replica index $\alpha$. Physically, $\mathcal{K}_1(E)$ corresponds to the disorder-averaged LDOS: 
\begin{equation}
\mathcal{K}_1(E)=\frac{\langle\rho(E)\rangle_{\rm dis}}{\rho_0}, \quad \rho(E) =  -\frac{1}{\pi}\im  G^+_E(\bm{r}, \bm{r}) .
\end{equation}
Here $G^+_E(\bm{r}, \bm{r^\prime})$ is  the single-particle Green's function for a given realization of disorder. 
The quantity $\rho_0$ stands for the density of states (including spin) at the energy $E \sim 1/\tau$. We note that the energy $E$ is counted from the Fermi level. Since $\mathcal{K}_1(E)$ represents the disorder-averaged density of states this is no surprise that it is pure scaling operator under the action of the renormalization group. 

The local operator without derivatives which involves two $Q$ matrices can be written as follows
\begin{equation}
\mathcal{K}_2(E_1,E_2) = \frac{1}{64}\sum_{p_1,p_2=\pm} p_1p_2 \mathcal{P}_2^{\alpha_1\alpha_2;p_1p_2}(E_1,E_2) .
 \label{eq:K2}
 \end{equation}
Here the correlation function $\mathcal{P}_2^{\alpha_1\alpha_2;p_1p_2}(E_1,E_2)$ can be obtained from
the Matsubara correlation function
 \begin{gather}
 \mathcal{P}_2^{\alpha_1\alpha_2}(i\varepsilon_{n_1},i\varepsilon_{n_2}) =\bigl  \langle \spp Q_{n_1n_1}^{\alpha_1\alpha_1}(\bm{r}) \spp Q_{n_2n_2}^{\alpha_2\alpha_2}(\bm{r}) \bigr \rangle \notag \\
   +\mu_2  \bigl \langle \spp \bigl [Q_{n_1n_2}^{\alpha_1\alpha_2}(\bm{r}) Q_{n_2n_1}^{\alpha_2\alpha_1}(\bm{r}) \bigr ] \bigr \rangle 
 \label{eq:P2M}
 \end{gather}
by standard analytic continuation to the real frequencies:  $\varepsilon_{n_1} \to E_1+i p_10^+ $ and $\varepsilon_{n_2} \to E_2+i p_2 0^+$. Here $\alpha_1$ and $\alpha_2$  are two different fixed replica indices. 
The operator $\mathcal{K}_2(E_1,E_2)$ is parametrized by a real number $\mu_2$. For $\mu_2=-2$ this operator describes the two-point correlation function of LDOS averaged over disorder realizations:
\begin{equation}
\mathcal{K}_2(E_1,E_2)\Bigl |_{\mu_2=-2} = \rho_0^{-2}\bigl  \langle \rho(E_1,\bm{r}) \rho(E_2,\bm{r})\bigr \rangle_{\rm dis} .
\label{eq:k2:LDOS}
\end{equation}
Recently,  it was shown within the two-loop approximation to the Finkel'stein NLSM that the operator $\mathcal{K}_2(E_1,E_2)$ in the case of $\mu_2=-2$ is the pure scaling operator \cite{Burmistrov2015a}. 

A general operator which involves the number $q$ of matrix fields $Q$ can be defined as 
\begin{gather}
\mathcal{K}_q(E_1,\dots,E_q) = \frac{1}{2^{3q}} \sum_{p_1,\dots p_q =\pm} \left ( \prod\limits _{j=1}^q p_j\right )
\notag \\
\times
\mathcal{P}_q^{\alpha_1,\dots,\alpha_q;p_1,\dots,p_q}(E_1,\dots,E_q)  .
 \label{eq:Kq}
 \end{gather}
Here $\mathcal{P}_q^{\alpha_1,\dots,\alpha_q;p_1,\dots,p_q}(E_1,\dots,E_q)$ can be read from the imaginary time correlation function
\begin{gather}
 \mathcal{P}_q^{\alpha_1,\dots,\alpha_q}(i\varepsilon_{n_{1}},\dots,i\varepsilon_{n_{q}}) = 
 \sum\limits_{\{k_1,\dots,k_q\}} \mu_{k_1,\dots, k_q} \bigl \langle  \mathcal{A}_{k_1,\dots, k_q} \bigr \rangle ,
 \notag \\
 \mathcal{A}_{k_1,\dots, k_q}=
 \prod\limits_{r=k_1}^{k_q}  
   \spp \bigl [Q_{n_{j_1}n_{j_2}}^{\alpha_{j_1} \alpha_{j_2}} Q_{n_{j_2}n_{j_3}}^{\alpha_{j_2}\alpha_{j_3}}\dots Q_{n_{j_r}n_{j_1}}^{\alpha_{j_r}\alpha_{j_1}}\bigr ]   ,
   \label{eq:PqM}
 \end{gather}
after the analytic continuation to the real frequencies:  $\varepsilon_{n_j} \to E_j+i p_j 0^+$. 
The sum in Eq. \eqref{eq:PqM} is performed over partitions of $q$, i.e. over all sets of positive integer numbers $\{k_1,\dots,k_q\}$ which satisfy the following conditions: $k_1+k_2+\dots k_q =q$ and $k_1\geq k_2 \geq \dots \geq k_q>0$. 
There is no summations over replica indices which are assumed to be all different, $\alpha_j \neq \alpha_k$ if $j\neq k$ for $j,k = 1, \dots, q$. In what follows we choose such normalization that $\mu_{1,1,\dots,1} = 1$.

In the absence of interactions, $\Gamma_j=0$, the NLSM action reduces to the first line in Eq. \eqref{eq:NLSM}. Since the energy of diffusive modes conserves without interactions, one can project $Q$ matrix to the $2\times 2$ subspace of given positive and negative Matsubara frequencies. 
In this way the group $G$ reduces to $\tilde{G}={\rm Sp}(8 N_r)$ and the effective action becomes $K$-invariant, i.e. invariant under rotations $Q \to U^{-1} Q U$ with $U \in \tilde{K} = {\rm Sp}(4N_r) \times {\rm Sp}(4N_r)$. Then operators $\mathcal{K}_q$ can be averaged over $U$ rotations and resulting $K$-invariant operators can be classified with respect to irreducible representations of $\tilde{G}$. Each irreducible representation will contain single $K$-invariant pure scaling operator \cite{Wegner1986,Gruzberg}. We note that (i) the above classification can be done for an arbitrary  number of replica $N_r$ and (ii) one can work with non $K$-invariant operators as well.

In the presence of interactions, application of the scheme described above is complicated by the following reasons. At first, the Finkel'stein NLSM action is not $K$-invariant. Secondly, the classification of $\mathcal{K}_q$ operators with respect to irreducible representations of the group $G$ is far from being obvious due to the limit $N_m\to \infty$. Because of these circumstances we employ two-loop renormalization procedure in order to fix sets of the coefficients $\mu_{k_1,\dots,k_q}$ which correspond to pure scaling operators.

\section{Two-loop renormalization of $\mathcal{K}_q$ \label{Sec:2loop}}

In order to find a set of proper coefficients $\mu_{k_1,\dots,k_q}$ which correspond to a pure scaling operator $\mathcal{K}_q$ we perform two-loop renormalization of the operator $\mathcal{K}_q$. 
As well known, in such perturbative (in $1/g$) treatment one encounters divergencies. For the purpose of regularization of the quantum theory, we add the mass term to the NLSM  action \eqref{eq:NLSM}:
\begin{equation}
S \to S_h = S + \frac{g h^2}{8} \int d \bm{r}\Tr \Lambda Q .
\label{SsGenFull}
\end{equation}
In addition, we will use the dimensional regularization scheme and start to work in $d=2+\epsilon$ dimensions.

For the perturbative treatment (in $1/g$) of the action \eqref{eq:NLSM}
one needs to resolve the nonlinear constraint $Q^2(\bm{r})=1$. For this purpose, we  
use the square-root parametrization
\begin{gather}
Q = W +\Lambda \sqrt{1-W^2}\ , \qquad W= \begin{pmatrix}
0 & w\\
\bar{w} & 0
\end{pmatrix} .
\label{eq:Q-W}
\end{gather}
In what follows we adopt the following notations: $W_{nm} = w_{nm}$ and $W_{mn} = \bar{w}_{mn}$ where $n\geqslant 0$ and $m<0$. The two blocks of matrix $W$  are related as
\begin{gather}
\bar{w} = -C w^T C,\qquad w = - C w^* C .
\end{gather}

Expanding the action $S_h$  in $W$ to the second order, we find the following propagators for the diffusive modes. For $r=0,3$ and $j=0,1,2,3$ they become
\begin{gather}
\Bigl \langle [w_{rj}(\bm{p})]^{\alpha_1\beta_1}_{n_1m_1} [\bar{w}_{rj}(-\bm{p})]^{\beta_2\alpha_2}_{m_2n_2} \Bigr \rangle =  \frac{2}{g} \delta^{\alpha_1\alpha_2} \delta^{\beta_1\beta_2}\delta_{n_{12},m_{12}}\notag \\
\times  \mathcal{D}_p(i\Omega_{nm}^\varepsilon)\Bigl [\delta_{n_1n_2} - \frac{32 \pi T \Gamma_j}{g}\delta^{\alpha_1\beta_1}  \mathcal{D}_p^{(j)}(i\Omega_{nm}^\varepsilon) \Bigr ] ,
\label{eq:prop:PH}
\end{gather}
where $n_{12}= n_1-n_2$, $m_{12}=m_1-m_2$, $\Omega_{nm}^\varepsilon = \varepsilon_{n_1}-\varepsilon_{m_1}\equiv \varepsilon_{n_2}-\varepsilon_{m_2}$, and $w_{rj} = \spp[w t_{rj}]/4$. The propagator \eqref{eq:prop:PH} involves a standard diffuson
\begin{equation}
\mathcal{D}^{-1}_p(i\omega_n) =p^2+h^2+{16 Z_\omega |\omega_n|}/{g} 
\label{eq:prop:Free}
\end{equation}
and diffusons renormalized by interaction in the singlet  ($\mathcal{D}_p^{(0)}(\omega) \equiv \mathcal{D}_p^{s}(\omega)$) and triplet  ($\mathcal{D}_p^{(1)}(\omega)=\mathcal{D}_p^{(2)}(\omega)=\mathcal{D}_p^{(3)}(\omega) \equiv \mathcal{D}_p^{t}(\omega)$)  channels:
\begin{equation}
[\mathcal{D}^{s/t}_p(i\omega_n)]^{-1}  =  p^2+h^2+{16 (Z_\omega+\Gamma_{s/t}) |\omega_n|}/{g} .
 \label{eq:prop:Int}
\end{equation}
The propagators of modes with $r=1,2$ and $j=0,1,2,3$ (cooperons) coincide with standard diffusons in the absence of interaction in the Cooper channel:
\begin{align}
\Bigl \langle [w_{rj}(\bm{p})]^{\alpha_1\beta_1}_{n_1m_1} [\bar{w}_{rj}(-\bm{p})]^{\beta_2\alpha_2}_{m_2n_2} \Bigr \rangle & =  \frac{2}{g} \delta^{\alpha_1\alpha_2} \delta^{\beta_1\beta_2} \delta_{n_1n_2}\notag \\
&\times
 \delta_{m_1m_2}\mathcal{D}_p(i\Omega_{nm}^\varepsilon) .
 \label{eq:prop:PPT}
\end{align}

At first, we remind known results of the one-loop renormalization. The disorder-averaged LDOS 
at $T=E=0$ can be written as \cite{Castellani1984}:
\begin{equation}
	\frac{\langle \rho(E=0)\rangle_{\rm dis}}{\rho_0}= \sqrt{Z} , \quad Z = 1 - \frac{h^\epsilon t}{\epsilon}\sum_{j=0}^3  \ln(1+\gamma_j) .
\label{eq:dos}
\end{equation}
Here we introduce $\gamma_j=\Gamma_j/Z_\omega$ and dimensionless resistivity $t =  8 \Omega_d/g$ where $\Omega_d= 1/[2^d\pi^{d/2}\Gamma(d/2)]$ ($t=2/(\pi g)$ in $d=2$). The renormalized conductance is given as \cite{Altshuler1979b,*Altshuler1980,Finkelstein1983b,*Finkelstein1984,Castellani1984} 
\begin{gather}
g^\prime = g \Bigl [1+\frac{a_1 t\, h^\epsilon}{\epsilon}\Bigr ], \quad
 a_1 =1+\sum_{j=0}^3 f(\gamma_j) ,
\label{eqS1}
\end{gather}
where
\begin{gather}
f(\gamma) = 1-(1+1/\gamma)\ln(1+\gamma)  .
\end{gather}
It will also important to take into account renormalization of  the momentum scale $h$ \cite{Baranov1999b}. The corresponding renormalized momentum scale $h^\prime$ is related with $h$  as follows $g^\prime h^{\prime 2} = g h^2 Z^{1/2}$. Within the one-loop approximation, one can find \cite{Baranov1999b} 
\begin{equation}
h^\prime = h \Biggl \{  1 - \frac{t\,  h^\epsilon}{2\epsilon} \Bigl [ 1+ \sum_{j=0}^3\bigl [ f(\gamma_j)+\frac{1}{2}\ln(1+\gamma_j) \bigr ] \Bigr ] \Biggr \}  .
\label{eqhren}
\end{equation}

In order to find renormalization of the operator $\mathcal{K}_q$ within two-loop approximation it is convenient to consider its irreducible part $\widetilde{\mathcal{K}}_q$ which can be obtained from definition \eqref{eq:PqM} by substitution $Q\to  Q - \langle Q \rangle$. From here and onwards $\langle \dots \rangle$ stands for the average with respect to the action $S_h$. This trick makes it possible to use only one-loop expression \eqref{eq:dos} for $Z$. Additional reason to work with irreducible operator $\widetilde{\mathcal{K}}_q$ is the fact that two-loop contribution to it vanishes for $q\geqslant 5$. Since the two-loop computation is standard and very close to the one presented in Ref. \cite{Burmistrov2015a}, here we present final result. Necessary technical details are given in Appendix \ref{App1}. Within the two-loop approximation a reducible operator $\mathcal{K}_q$ can be written in the following form at $T=0$
\begin{equation}
\mathcal{K}_q(E_1=0,\dots,E_q=0) = Z^{q/2}\ m^\prime_q,\qquad q\geqslant 2 .
\end{equation}
Here 
\begin{equation}
m^\prime_q = m_q \Bigl [ 1 +  \frac{b_1 t \, h^{\prime \epsilon}}{\epsilon}+ \frac{t^2 h^{\prime 2\epsilon}}{\epsilon^2} \Bigl (b_2+\epsilon b_3\Bigr )\Bigr ] ,
\label{eq:m2l:ren}
\end{equation}
where $m_q=1$ due to normalization convention for operators $\mathcal{K}_q$. The coefficients $b_{1,2,3}$ depend on the interaction parameters $\gamma_j$ and coefficients $\mu_{k_1,\dots,k_q}$:
\begin{gather}
b_1 = \mu_{2,1,1,\dots,1}, \qquad b_3 = \mu_{2,1,1,\dots,1} \sum_{j=0}^3 c(\gamma_j)/4 ,
\notag \\
b_2 = q(q-1)/2 + 3 \mu_{3,1,1,\dots,1}/2 + \mu_{2,2,1,\dots,1} - \mu_{2,1,1,\dots,1}\notag \\
 - 
\mu_{2,1,1,\dots,1} \sum_{j=0}^3 f(\gamma_j)/2 .
\label{eq:b123}
\end{gather}
The function
\begin{gather}
c(\gamma) = 2 +\frac{2+\gamma}{\gamma} \liq(-\gamma) + \frac{1+\gamma}{2\gamma} \ln^2(1+\gamma) 
\label{eq:def:c}
\end{gather}
involves the polylogarithm $\liq(x) = \sum_{k=1}^\infty x^k/k^2$. We note that  $\mu_{3,1,1,\dots,1}=0$ for $q=2$ and $\mu_{2,2,1,\dots,1}=0$ for $q=2, 3$. Applying the minimal subtraction scheme (see e.g. \cite{Amit-book}) to Eq. \eqref{eq:m2l:ren}, we can obtain the anomalous dimension of $m_q^\prime$. In order to have it finite in the limit $\epsilon\to 0$ the condition $b_2=b_1(b_1-a_1)/2$ should be satisfied. As follows from
Eqs. \eqref{eq:b123} this condition is equivalent to the following equation for the coefficients $\mu_{k_1,\dots,k_q}$:
\begin{equation}
 \mu_{2,1,\dots,1}^2+\mu_{2,1,\dots,1} = 3\mu_{3,1,\dots,1} +2\mu_{2,2,\dots,1} +q(q-1) .
\label{eq:cons-cond}
\end{equation}
Since the pure scaling operator $\mathcal{K}_q$ should have a well-defined anomalous dimension, Eq. \eqref{eq:cons-cond} should be satisfied by a set of its coefficients $\mu_{k_1,\dots,k_q}$. Moreover, since 
Eq. \eqref{eq:cons-cond} is nonlinear for $\mu_{2,1,\dots,1}$, in general, it cannot be satisfied with coefficients $\mu_{k_1,\dots,k_q}$ which correspond to linear combinations of two pure scaling operators. Surprisingly, Eq. \eqref{eq:cons-cond} is independent of interaction parameters $\gamma_s$ and $\gamma_t$. This means that Eq. \eqref{eq:cons-cond} is satisfied by the sets $\{\mu_{k_1,\dots,k_q}\}$ which determine pure scaling operators in the noninteracting theory \cite{Wegner1986,Wegner1987a,*Wegner1987b}. For $q=2,3$ and $4$ we list the sets $\{\mu_{k_1,\dots,k_q}\}$ corresponding to pure scaling operators in Table \ref{Tab}. As one can check they satisfy Eq. \eqref{eq:cons-cond}. We mention that for $q=2$ Eq. \eqref{eq:cons-cond} allows to determine possible values of the coefficient $\mu_2$ (equal to -2 and 1) and, therefore, allows us to find all possible pure scaling operators with $q=2$. In order to determine all coefficients in the set $\{\mu_{k_1,\dots,k_q}\}$ for $q\geqslant 3$, one needs to analyze next orders in the loop expansion. In the next section, we present arguments based on the background field renormalization method why we expect that the sets $\{\mu_{k_1,\dots,k_q}\}$ which determine the pure scaling operators in the noninteracting problem do determine the pure scaling operators in the interacting problem as well.

Provided Eq. \eqref{eq:cons-cond} is satisfied, the anomalous dimension of the corresponding pure scaling operator in the two-loop approximation is given as
\begin{equation}
-\frac{d \ln m^\prime_q}{d y} = \zeta^{(\mu)}_q =   \mu_{2,1,1,\dots,1}\Bigl [ t +\frac{t^2}{2} \sum_{j=0}^3 c(\gamma_j)\Bigr ]  + O(t^3) .
\label{eqm2RG}
\end{equation}
Here $y=\ln 1/h^\prime$. We emphasize that within the two-loop approximation the anomalous dimension of the pure scaling operator $\zeta^{(\mu)}_q$ is fully determined by the single coefficient  $\mu_{2,1,1,\dots,1}$. The critical exponent $\Delta_q^{(\mu)}$ for the pure scaling operator $\mathcal{K}_q^{(\mu)}$ is equal to the corresponding anomalous dimension $\zeta^{(\mu)}_q$ at the fixed point. 

We remind that similar situation is known for the noninteracting problem where the anomalous dimension of pure scaling operators are known upto the forth loop order \cite{Wegner1986,Wegner1987a,*Wegner1987b}:
\begin{equation}
\zeta^{(\mu)}_{q,{\rm n}} =   \mu_{2,1,1,\dots,1} t   + c_3 \zeta(3) t^4 + O(t^5) .
\label{eqm2RG-ni}
\end{equation}
Here $\zeta(z)$ denotes the Riemann zeta function and the coefficients $c_3$ computed in Refs. \cite{Wegner1986,Wegner1987a,*Wegner1987b} are listed in Table \ref{Tab} for $q=2,3$ and 4. Since $c(0)=0$ our result \eqref{eqm2RG} is in agreement with the noninteracting result \eqref{eqm2RG-ni}.

\begin{table}
\caption{The coefficients $\mu_{k_1,\dots,k_q}$ and $c_3$ for the pure scaling operators with $q=2,3$ and $4$  taken from Refs. \cite{Wegner1986,Wegner1987a,*Wegner1987b} (see text).  \label{Tab}}
\begin{ruledtabular}
\begin{tabular}{cccccc}
\multicolumn{6}{c}{$q=2$} \\
\hline
$\mu_{1,1}$ & 1 & 1 & & & \\
$\mu_2$ & -2 & 1 & & & \\
$c_3$ &3/2 &-3/2 & & & \\
\hline
\multicolumn{6}{c}{$q=3$} \\
\hline
$\mu_{1,1,1}$ & 1 & 1 & 1 & & \\
$\mu_{2,1}$ & -6 & -1 & 3 & & \\
$\mu_{3}$ & -8 & -2 & 2 & & \\
$c_3$ & 21/2 & 3 & -12 & & \\
\hline
\multicolumn{6}{c}{$q=4$} \\
\hline
$\mu_{1,1,1,1}$ & 1 & 1 & 1 & 1 & 1\\
$\mu_{2,1,1}$ & -12 & -5 & -2 & 1& 6\\
$\mu_{2,2}$ & 12 & -2 & 7 & -2& 3\\
$\mu_{3,1}$ & 32 & 4 & -8 & -2 & 8\\
$\mu_{4}$ & -48 & 8 & 2 & -4 & 6\\
$c_3$ &39 &18 & 3/2& -3/2& -93/2\\
\end{tabular}
\end{ruledtabular}
\end{table}

\section{Background field renormalization \label{Sec:BFM}}

The results of the previous section indicate that the sets of coefficients $\{\mu_{k_1,\dots,k_q}\}$ which determine local pure scaling operators without derivatives for the Finkel'stein NLSM coincide, in fact, with ones known from the noninteracting theory. Below in this section we present additional arguments based on the background field method which support this observation.

In order to employ the background field renormalization method we split the matrix field $Q$ into fast $\hat{Q}$ and slow $Q_0=\mathcal{T}_0^{-1} \Lambda \mathcal{T}_0$ components:  $Q = \mathcal{T}_0^{-1} \hat{Q} \mathcal{T}_0$. The fast and slow modes are separated in the Matsubara frequency space by an energy scale $\mathcal{E}_\Lambda$ such that $2\pi T \ll \mathcal{E}_\Lambda \ll 2\pi T N_m$. In particular, the slow mode is trivial at high frequencies: $(\mathcal{T}_0)_{nm}^{\alpha\beta} = \delta_{nm} \delta^{\alpha\beta} t_{00}$ if $|\varepsilon_n| > \mathcal{E}_\Lambda$ or $|\varepsilon_m| > \mathcal{E}_\Lambda$. 

The first non-trivial example is the background field renormalization of the bilinear in $Q$ operators. 
For them there are two basis operators $\mathcal{A}_{1,1}$ and $\mathcal{A}_{2}$. In the background field method they transform as  
\begin{align}
\mathcal{A}_{1,1} & \to \bigl \langle \spp \bigl [ (\mathcal{T}_0^{-1})^{\alpha_1\alpha_1^\prime}_{nn^\prime} \hat{Q}_{n^\prime n^{\prime\prime}}^{\alpha^\prime_1\alpha_1^{\prime\prime}} (\mathcal{T}_0)^{\alpha_1^{\prime\prime}\alpha_1}_{n^{\prime\prime}n} \bigr  ] \notag \\
& \times \spp \bigl [ (\mathcal{T}_0^{-1})^{\alpha_2\alpha_2^\prime}_{mm^\prime} \hat{Q}_{m^\prime m^{\prime\prime}}^{\alpha^\prime_2\alpha_2^{\prime\prime}} (\mathcal{T}_0)^{\alpha_2^{\prime\prime}\alpha_2}_{m^{\prime\prime}m} \bigr ] \bigr\rangle_f, \notag \\
\mathcal{A}_{2}& \to \bigl \langle \spp \bigl [ (\mathcal{T}_0^{-1})^{\alpha_1\alpha_1^\prime}_{nn^\prime} \hat{Q}_{n^\prime n^{\prime\prime}}^{\alpha^\prime_1\alpha_1^{\prime\prime}} (\mathcal{T}_0)^{\alpha_1^{\prime\prime}\alpha_2}_{n^{\prime\prime}m} \notag \\
& \times (\mathcal{T}_0^{-1})^{\alpha_2\alpha_2^\prime}_{mm^\prime} \hat{Q}_{m^\prime m^{\prime\prime}}^{\alpha^\prime_2\alpha_2^{\prime\prime}} (\mathcal{T}_0)^{\alpha_2^{\prime\prime}\alpha_1}_{m^{\prime\prime}n} \bigr]  \bigr \rangle_f .
\label{eq:a11a2:app}
\end{align}
Here $\langle \dots \rangle_f$ denotes the average over fast modes $\hat Q$. We emphasize that in order the right hand sides of Eqs. \eqref{eq:a11a2:app} to result in the bilinear in $Q_0$ contributions the Matsubara frequency indices of matrices $\mathcal{T}_0$ and $\mathcal{T}_0^{-1}$ should be `small',
$\max\{|\varepsilon_{n^\prime}|, |\varepsilon_{n^{\prime\prime}}|, |\varepsilon_{m^\prime}|, |\varepsilon_{m^{\prime\prime}}|\} <\mathcal{E}_\Lambda$. We note that the separation on slow and fast modes is enough to perform in the operators only. The renormalization of the NLSM action and renormalization of these bilinear in $Q$ operators with small Matsubara frequencies are independent from each other. Therefore, in order to evaluate averages in the right hand side of Eq. \eqref{eq:a11a2:app} one needs to know the two-point correlation function of the fast fields, $\langle  \hat{Q}_{n^\prime n^{\prime\prime}}^{\alpha^\prime_1,\alpha_1^{\prime\prime}}   \hat{Q}_{m^\prime m^{\prime\prime}}^{\alpha^\prime_2,\alpha_2^{\prime\prime}} \rangle_f$, with `small' Matsubara frequencies only.  

We start from treating the two-point correlation function of the fast modes within one-loop approximation. 
Expanding the matrix field $\hat{Q}$ to the second order in $W$ (see Eq. \eqref{eq:Q-W}) and using 
Eqs. \eqref{eq:prop:PH} and \eqref{eq:prop:PPT}, we find 
\begin{gather}
\begin{pmatrix}
\mathcal{A}_2[Q]  \\
\mathcal{A}_{1,1}[Q] 
\end{pmatrix}  \to 
\bigl [1+2(Z^{1/2}-1)\bigr ]\begin{pmatrix}
\mathcal{A}_2[Q_0] \\
\mathcal{A}_{1,1}[Q_0]
\end{pmatrix} \notag \\
- \frac{t h^\epsilon}{\epsilon}
\mathcal{M}_2
\begin{pmatrix}
\mathcal{A}_2[Q_0] \\
\mathcal{A}_{1,1}[Q_0] 
\end{pmatrix}  ,
\label{eq:bgfr:1}
\end{gather} 
where the mixing matrix
\begin{equation}
\mathcal{M}_2=\begin{pmatrix}
1 & -1 \\
-2 & 0
\end{pmatrix} .
\end{equation}
The eigenvalues (with a minus sign) of the matrix $\mathcal{M}_2$ are equal to $\lambda_2^{(1)} = -2$ and $\lambda_2^{(2)} = 1$. The corresponding eigenoperators becomes
\begin{equation}
\mathcal{P}_{2}^{(-2)} = \mathcal{A}_{1,1}-2 \mathcal{A}_2, \qquad \mathcal{P}_{2}^{(1)} = \mathcal{A}_{1,1}+\mathcal{A}_2 .
\label{Q2}
\end{equation} 
This implies the values of $\mu_2$ equal to $-2$ and $1$ in accordance with Table \ref{Tab}. We emphasize that within one-loop approximation the mixing matrix $\mathcal{M}_2$ is independent of interaction parameters $\gamma_s$ and $\gamma_t$. They appear in the overall factor $[1+2(Z^{1/2}-1)]$ only.  

Next we argue why the mixing matrix $\mathcal{M}_2$ remains the same in all orders in $t$. We start from the correlation function $\langle  \hat{Q}_{n^\prime n^{\prime\prime}}^{\alpha^\prime_1\alpha_1^{\prime\prime}}   \hat{Q}_{m^\prime m^{\prime\prime}}^{\alpha^\prime_2\alpha_2^{\prime\prime}} \rangle_f$ with 
$n^\prime n^{\prime\prime} <0$ and $m^\prime m^{\prime\prime}<0$. Since in the case of $nm<0$ the expansion of $Q_{nm}$ in powers of $W$ in the square-root parametrization consists of the single term -- $W$, the correlation function has the same structure as the propagator $\langle  W_{n^\prime n^{\prime\prime}}^{\alpha^\prime_1\alpha_1^{\prime\prime}}  W_{m^\prime m^{\prime\prime}}^{\alpha^\prime_2\alpha_2^{\prime\prime}} \rangle$ but with the renormalized parameters. Within one-loop approximation this fact was verified explicitly for $\alpha^\prime_1\neq \alpha_1^{\prime\prime}$ in Ref. \cite{Burmistrov2015a}. In general, we can write  for $n^\prime n^{\prime\prime}<0$, $m^\prime m^{\prime\prime}<0$ and $n^\prime m^{\prime\prime}>0$  (cf. Eq. \eqref{eq:prop:PH})
\begin{gather}
\Bigl \langle [\hat{Q}_{rj}]_{n^\prime n^{\prime\prime}}^{\alpha^\prime_1\alpha_1^{\prime\prime}} [\hat{Q}_{rj}]_{m^\prime m^{\prime\prime}}^{\alpha^\prime_2\alpha_2^{\prime\prime}} \Bigr \rangle = Z \delta^{\alpha^\prime_1\alpha_1^{\prime\prime}} \delta^{\alpha^\prime_2\alpha_2^{\prime\prime}}\delta_{n^\prime - n^{\prime\prime}, m^{\prime\prime}-m^\prime}\notag \\
\times  B_1  \Bigl [\delta_{n^\prime m^{\prime\prime}} + T C_1 \delta^{\alpha^\prime_1\alpha_1^{\prime\prime}}   \Bigr ] , 
\label{eq:prop:PH-full-1}
\end{gather}
in the case $r=0,3$ and $j=0,1,2,3$,  and 
\begin{gather}
\Bigl \langle [\hat{Q}_{rj}]_{n^\prime n^{\prime\prime}}^{\alpha^\prime_1\alpha_1^{\prime\prime}} [\hat{Q}_{rj}]_{m^\prime m^{\prime\prime}}^{\alpha^\prime_2\alpha_2^{\prime\prime}} \Bigr \rangle = Z \tilde{B}_1\delta^{\alpha^\prime_1\alpha_1^{\prime\prime}} \delta^{\alpha^\prime_2\alpha_2^{\prime\prime}}
 \delta_{n^\prime m^{\prime\prime}} \delta_{n^{\prime\prime} m^\prime} 
\label{eq:prop:PP-full-1}
\end{gather}
in the case  $r=1,2$ and $j=0,1,2,3$. Here the functions $B_1$, and $C_1$ are diffusons with renormalized parameters $g$, $\Gamma_j$ and $Z_\omega$. The cooperon $\tilde{B}_1$ as well as mesoscopic diffusons $B_1$, and $C_1$ contain in their denominators the corresponding dephasing times induced by interaction and disorder. We consider length scales shorter than the dephasing length such that the renormalized cooperon $\tilde{B}_1$ becomes the same as the renormalized diffuson $B_1$. We stress that the second term in the right hand side of Eq. \eqref{eq:prop:PH-full-1} is proportional to the temperature and thus can be omitted at $T\to 0$.  The correlation function $\langle  \hat{Q}_{n^\prime n^{\prime\prime}}^{\alpha^\prime_1\alpha_1^{\prime\prime}}   \hat{Q}_{m^\prime m^{\prime\prime}}^{\alpha^\prime_2\alpha_2^{\prime\prime}} \rangle_f$ for 
$n^\prime n^{\prime\prime}<0$, $m^\prime m^{\prime\prime}<0$ and $n^\prime m^{\prime\prime}<0$ can be obtained from Eqs. \eqref{eq:prop:PH-full-1} and \eqref{eq:prop:PP-full-1} with the help of Eq. \eqref{TC2}.

Next we consider the correlation function $\langle  \hat{Q}_{n^\prime n^{\prime\prime}}^{\alpha^\prime_1\alpha_1^{\prime\prime}}   \hat{Q}_{m^\prime m^{\prime\prime}}^{\alpha^\prime_2\alpha_2^{\prime\prime}} \rangle_f$ with $n^\prime n^{\prime\prime}<0$ and $m^\prime m^{\prime\prime}>0$. In order this correlation function to be nonzero one has to take into account the non-Gaussian part of the action $S_h$ which is odd in $W$. The simplest contribution is due to the interaction part of the action $S_h$ which is of the third order in $W$. Evaluating the averages with the help of the Wick's theorem, one can see that the summation index $n$ in Eq. \eqref{eq:NLSM} becomes restricted to the region of small energies, $|\omega_n| <\mathcal{E}_\Lambda$. Thus, due to the presence of $T$ in the interaction part of the action \eqref{eq:NLSM}, this simplest contribution to the two-point correlation function with $n^\prime n^{\prime\prime}<0$ and $m^\prime m^{\prime\prime}>0$ becomes proportional to the temperature. This argument can be extended to contributions of the higher order in $W$. All in all, for $n^\prime n^{\prime\prime}<0$ and $m^\prime m^{\prime\prime}>0$ we find that  
\begin{equation}
\langle  \hat{Q}_{n^\prime n^{\prime\prime}}^{\alpha^\prime_1\alpha_1^{\prime\prime}}   \hat{Q}_{m^\prime m^{\prime\prime}}^{\alpha^\prime_2\alpha_2^{\prime\prime}} \rangle_f \propto T .
\label{eq:prop:+-++}
\end{equation}
Thus the two-point correlation function with $n^\prime n^{\prime\prime}<0$ and $m^\prime m^{\prime\prime}>0$ does not contribute at $T=0$. 

Next we consider $\langle  \hat{Q}_{n^\prime n^{\prime\prime}}^{\alpha^\prime_1\alpha_1^{\prime\prime}}   \hat{Q}_{m^\prime m^{\prime\prime}}^{\alpha^\prime_2\alpha_2^{\prime\prime}} \rangle_f$ with
$n^\prime n^{\prime\prime}>0$, $m^\prime m^{\prime\prime}>0$, and $n^\prime m^\prime>0$.
In this case the lowest order contribution comes from the expansion of each fast field $\hat Q$ to the second order in $W$. The evaluation of averages with the help of the Wick's theorem and Eqs. \eqref{eq:prop:PH} and \eqref{eq:prop:PPT} suggests the following general result at $T=0$:
\begin{gather}
\langle  [\hat{Q}_{rj}]_{n^\prime n^{\prime\prime}}^{\alpha^\prime_1\alpha_1^{\prime\prime}}   [\hat{Q}_{rj}]_{m^\prime m^{\prime\prime}}^{\alpha^\prime_2\alpha_2^{\prime\prime}} \rangle_f =
 Z \delta_{r0}\delta_{j0}
 \delta_{n^\prime n^{\prime\prime}}\delta_{m^\prime m^{\prime\prime}}  \delta^{\alpha^\prime_1\alpha_1^{\prime\prime}} \delta^{\alpha^\prime_2\alpha_2^{\prime\prime}} \notag \\
 + B_{2} \delta_{n^\prime m^{\prime\prime}} \delta_{m^\prime n^{\prime\prime}}  \delta^{\alpha^\prime_1\alpha_2^{\prime\prime}} 
 \delta^{\alpha^\prime_2\alpha_1^{\prime\prime}} 
 - \frac{\tilde{B}_2}{4} \spp [t_{rj} C t_{rj}^{\rm T} C]\notag \\
 \times \delta_{n^\prime m^\prime}\delta_{n^{\prime\prime}m^{\prime\prime}} \delta^{\alpha^\prime_1\alpha^\prime_2} \delta^{\alpha_1^{\prime\prime}\alpha_2^{\prime\prime}} .
 \label{eq:prop:++++}
\end{gather}
We note that the following relation, $\tilde{B}_2 \equiv B_2$, holds due to the charge-conjugation constraint \eqref{TC2}.

At last, we discuss $\langle  \hat{Q}_{n^\prime n^{\prime\prime}}^{\alpha^\prime_1\alpha_1^{\prime\prime}}   \hat{Q}_{m^\prime m^{\prime\prime}}^{\alpha^\prime_2\alpha_2^{\prime\prime}} \rangle_f$ with $n^\prime n^{\prime\prime}>0$, $m^\prime m^{\prime\prime}>0$, and $n^\prime m^\prime<0$. As in the previous case, the lowest order contribution is given by expansion of $\hat Q$ to the second order in $W$. Evaluation
the averages with the help of Eqs. \eqref{eq:prop:PH} -- \eqref{eq:prop:PPT} and the Wick's theorem suggests the following result for $T=0$:
\begin{gather}
\langle  \hat{Q}_{n^\prime n^{\prime\prime}}^{\alpha^\prime_1,\alpha_1^{\prime\prime}}   \hat{Q}_{m^\prime m^{\prime\prime}}^{\alpha^\prime_2\alpha_2^{\prime\prime}} \rangle_f =
  Z B_3 \Lambda_{n^\prime n^{\prime\prime}}^{\alpha^\prime_1\alpha_1^{\prime\prime}} \Lambda_{m^\prime m^{\prime\prime}}^{\alpha^\prime_2\alpha_2^{\prime\prime}}  .
 \label{eq:prop:++--}
 \end{gather}
We emphasize that such simple form of the two-point correlation function of fast fields, Eqs. \eqref{eq:prop:PP-full-1} -- \eqref{eq:prop:++--}, is due to (i) zero-temperature limit and (ii) smallness of frequency indices, $\max\{|\varepsilon_{n^\prime}|, |\varepsilon_{n^{\prime\prime}}|, |\varepsilon_{m^\prime}|, |\varepsilon_{m^{\prime\prime}}|\} <\mathcal{E}_\Lambda$.
 
Using Eqs. \eqref{eq:prop:PP-full-1} -- \eqref{eq:prop:++--}, we  perform averaging over the fast fields in 
Eqs. \eqref{eq:a11a2:app} and find
\begin{gather}
\begin{pmatrix}
\mathcal{A}_2[Q]  \\
\mathcal{A}_{1,1}[Q] 
\end{pmatrix}  \to 
Z(1+B_3)/2\begin{pmatrix}
\mathcal{A}_2[Q_0] \\
\mathcal{A}_{1,1}[Q_0]
\end{pmatrix}  \notag \\
+2 Z (B_1-B_2)
\mathcal{M}_2
\begin{pmatrix}
\mathcal{A}_2[Q_0] \\
\mathcal{A}_{1,1}[Q_0] 
\end{pmatrix}  .
\label{eq:bgfr:12}
\end{gather}
Comparing Eq. \eqref{eq:bgfr:12} with Eq. \eqref{eq:bgfr:1}, one can see that Eq. \eqref{Q2} holds beyond one-loop approximation and, in general, the pure scaling operators bilinear in $Q$  correspond to $\mu_2=-2$ and $\mu_2=1$.

The background field renormalization method for local operators of higher order in $Q$ (for $q=3$ and $4$) within one-loop approximation for the fast fields is presented in Appendix \ref{App2}. Using arguments similar to given above one can extend the one-loop results of Appendix \ref{App2} to all orders in $t$.

\section{Discussions and conclusions \label{Sec:Sum}}

The results of the previous sections directly imply that the Finkel'stein NLSM supports a large class of pure scaling operators which can be constructed as a straightforward generalization of pure scaling operators known for the noninteracting NLSM. We emphasize that these operators, Eq. \eqref{eq:PqM}, are not invariant under rotations in the Matsubara and replica spaces. In striking contrast with the noninteracting case, the set of local pure scaling operators without derivatives in the Finkel'stein NLSM is not exhausted by these non-gauge invariant operators. For example, it is known \cite{Baranov1999a} that the operator in the second line of the NLSM  action \eqref{eq:NLSM} can be used to construct the gauge-invariant operator. In the Finkel'stein NLSM there exist local gauge invariant operators which involve three and four $Q$ matrices \cite{PruiskenVoropaev}. We note that in the case of such operators the background field renormalization of the NLSM action interferes with the renormalization of the operators \cite{Baranov1999b}.
The detailed discussion of such gauge-invariant operators is beyond the scope of the present paper. 

As we have already mentioned, the operator $\mathcal{K}_2^{(-2)}$ can be expressed as the second moment of the LDOS, see Eq. \eqref{eq:k2:LDOS}. Alternatively, the same operator can be written in terms of the following two-point correlation function of single-particle Green's function:
\begin{align}
\mathcal{K}_2^{(-2)} & \propto \bigl  \langle \im G^+_E(\bm{r}, \bm{r}) \im G^+_E(\bm{r^\prime}, \bm{r^\prime}) \notag \\
& +2  \im G^+_E(\bm{r^\prime},\bm{r}) \im G^+_E(\bm{r}, \bm{r^\prime})  \bigr \rangle_{\rm dis}  .
\label{eq:K2-2:GG}
\end{align}
Here we assume the distance between two points $\bm{r}$ and $\bm{r^\prime}$ satisfies the following condition: 
\begin{equation}
\lambda_F \ll |\bm{r}-\bm{r^\prime}| \ll l, 
\label{eq:cond:dist}
\end{equation}
where $\lambda_F$ and $l$ stand for the Fermi length and elastic mean free path, respectively.  The other pure scaling operator  bilinear in $Q$ can be expressed via single-particle Green's function in a similar way: 
\begin{align}
\mathcal{K}_2^{(1)} & \propto \bigl \langle \im G^+_E(\bm{r}, \bm{r}) \im G^+_E(\bm{r^\prime}, \bm{r^\prime})
\notag \\
&  -  \im G^+_E(\bm{r^\prime},\bm{r}) \im G^+_E(\bm{r}, \bm{r^\prime})  \bigr \rangle_{\rm dis} ,
\end{align}
where the points $\bm{r}$ and $\bm{r^\prime}$ satisfy condition \eqref{eq:cond:dist}. Alternatively, pure scaling operator $\mathcal{K}_2^{(1)}$ can be written in terms of the spatial correlation function of LDOS:
\begin{align}
\mathcal{K}_2^{(1)}  \propto \bigl \langle 3 \rho(E,\bm{r}) \rho(E,\bm{r^\prime})
 -  \rho(E,\bm{r}) \rho(E, \bm{r})  \bigr \rangle_{\rm dis} ,
\end{align}
where again the distance $|\bm{r} - \bm{r^\prime}|$ is restricted by the inequality \eqref{eq:cond:dist}. 
In general, all pure scaling operators $\mathcal{K}_q^{(\mu)}$ can be expressed in terms of disorder-averaged spatial correlation functions of single-particle Green's functions or, alternatively, LDOS. This means that the scaling behavior of pure scaling operators can be extracted from accurate analysis of data obtained from scanning tunneling microscopy. 

In this paper we consider the system with preserved time-reversal and spin-rotational symmetries. Similar results and conclusions can be obtained for the other cases in which time-reversal or/and spin-rotational symmetries are broken \cite{RB2}. In the case of broken spin-rotational but preserved time-reversal symmetries our results allow to predict the behavior of LDOS in scanning tunneling spectroscopy maps at the surface of three dimensional topological insulator.

In summary, we demonstrated that the local pure scaling operators without derivatives in the Finkel'stein nonlinear sigma model can be constructed by straightforward generalization of corresponding operators for the noninteracting case. These pure scaling operators describe mesoscopic fluctuations of the single-particle Green's function and the local density of states in the presence of electron-electron interaction.  We determined anomalous dimensions of such pure scaling operators in the interacting theory within the two-loop approximation. Similar to  the noninteracting case, approximately half of these pure scaling operators demonstrate multifractal behavior.

\begin{acknowledgements}
We thank I. Gornyi, I. Gruzberg, D. Lyubshin, and A. Mirlin for useful discussions. The work was partially supported by the Russian Foundation for Basic Research under the Grant No. 14-02-00033, Russian President Grant No. MD-5620.2016.2, Russian President Scientific Schools Grant NSh-10129.2016.2 and the Ministry of Education and Science of the Russian Federation under the Grant No. 14.Y26.31.0007.
\end{acknowledgements}

\appendix

\section{Details of the two-loop renormalization of $\mathcal{K}_q$ \label{App1}}

In this Appendix we present details of two-loop renormalization of operators $\mathcal{K}_q$. As we discussed in the main text it is convenient to work with irreducible operators $\widetilde{\mathcal{K}}_q$. The irreducible operators corresponding to $\mathcal{P}_q$ will be denoted as   $\widetilde{\mathcal{P}}_q$.

\begin{widetext}	
\subsection{Bilinear in $Q$ operators}
	
We start from operators $\mathcal{K}_2$ which are quadratic in $Q$. In the one-loop approximation, i.e. in the first order in $1/g$, there is nonzero contribution to $\widetilde{\mathcal{P}}^{\alpha_1\alpha_2}_2(i\varepsilon_{n_1},i\varepsilon_{n_2})$ for $n_1 n_2 <0$ only. We find for $n>0$ and $m<0$
	\begin{gather}
		\widetilde{\mathcal{P}}_2^{\alpha_1\alpha_2;(1)}(i\varepsilon_{n},i\varepsilon_{m}) =\mu_2 \langle \spp[ w_{n m}^{\alpha_1 \alpha_2}\overline{w}_{m n}^{\alpha_2 \alpha_1}]\rangle 
		= \frac{128\mu_2}{g}\int_q \mathcal{D}_q(i\Omega^\varepsilon_{nm})  .
	\end{gather}
Here we introduce $\int_q \equiv \int d^d\bm{q}/(2\pi)^d$. After the analytic continuation to real frequencies and setting afterwards $T=E_1=E_2= 0$ we find the following one-loop result
	\begin{equation}
		\widetilde{\mathcal{K}}^{(1)}_2 = \mu_2  \frac{h^\epsilon t}{\epsilon}  .
		\label{eq:K2:1-app}
	\end{equation}
					
In the second order in $1/g$ the nonzero contributions to $\widetilde{\mathcal{P}}^{\alpha_1\alpha_2}_2(i\varepsilon_{n_1},i\varepsilon_{n_2})$ exist for both with $n_1n_2>0$ and $n_1 n_2<0$. In the case $n_1>0$ and $n_2>0$, we can write
	\begin{equation}
	\widetilde{\mathcal{P}}_2^{\alpha_1\alpha_2;(2)}(i\varepsilon_{n_1},i\varepsilon_{n_2}) = \frac{\mu_2}{4} \sum_{m_1m_2}\sum_{\beta_1\beta_2} \bigl \langle 
	\spp[w^{\alpha_1\beta_1}_{n_1m_1}\bar{w}^{\beta_1\alpha_2}_{m_1n_2} 
	w^{\alpha_2\beta_2}_{n_2m_2}\bar{w}^{\beta_2\alpha_1}_{m_2n_1} ]
	\bigr \rangle  .
	\end{equation}
	Using Wick's theorem, we obtain
	\begin{equation}
	\widetilde{\mathcal{P}}_2^{\alpha_1\alpha_2;(2)}(i\varepsilon_{n_1},i\varepsilon_{n_2}) = -\mu_2
	\left (\frac{64}{g}\right )^2 \sum_{j=0}^{3}\frac{\pi T \Gamma_j}{g} \sum_{\omega_n>\varepsilon_{n_2}>0} \int_{q,p} \mathcal{D}_q(i\omega_n+i\Omega_{12})\mathcal{D}_p(i\omega_n)\mathcal{D}_p^{(j)}(i\omega_n)+(\varepsilon_1 \leftrightarrow \varepsilon_2).
	\label{eq2loopK2_P2++1}
	\end{equation} 
	After the analytic continuation to the real energies, $i \varepsilon_{n_{1,2}} \to E_{1,2} + i 0^+$,  we obtain
	\begin{equation}
	\widetilde{\mathcal{P}}_2^{\alpha_1\alpha_2;++;(2)}(E_1,E_2)  = 
	-\mu_2 \left (\frac{32}{g}\right )^2 \sum_{j=0}^{3}\frac{\Gamma_j}{i g} \int_{-\infty}^\infty d\omega \mathcal{F}_{\omega-E_2}\int_{q,p}
	\mathcal{D}^R_q(\omega+E_1-E_2) \mathcal{D}^R_p(\omega)\mathcal{D}^{R,(j)}_p(\omega)+(E_1\leftrightarrow E_2) ,
	\label{eq2loopK2_P2++11}
	\end{equation} 
	where $\mathcal{F}_\omega=\tanh[\omega/(2T)]$ stands for the equilibrium fermionic distribution function and $\mathcal{D}^R_p(\omega)$ denotes the retarded propagator corresponding to $\mathcal{D}_p(i\omega)$.	Again setting $E_1=E_2=T=0$ and using the Feynman's trick
	\begin{equation}
	\frac{1}{ABC}=\int_{0}^{1}\prod_{i=1}^{3}dx_i \delta\left (\sum_{i}^{}x_i-1\right )\frac{2}{(Ax_1+Bx_2+Cx_3)^3}
	\end{equation} 
	which allows us to perform integrations over momenta $\bm{q}$ and $\bm{p}$, we obtain
	\begin{equation}
	\widetilde{\mathcal{P}}_2^{\alpha_1\alpha_2;++;(2)}=\left (\frac{16}{g}\right )^2\sum_{j=0}^{3}\frac{\gamma_j  A_\epsilon\Gamma(1-\epsilon)h^{2\epsilon}}{\epsilon\Gamma^2(1-\epsilon/2)}
	\int_{0}^{1}\prod_{i=1}^{3}dx_i \delta\left (\sum_{i}^{}x_i-1\right )\frac{x_2^{-1-\epsilon/2}(1-x_2)^{-1-\epsilon/2}}{1+\gamma_j x_3} .
	\end{equation}
	Here $A_\epsilon=\Omega_d^2\Gamma^2(1-\epsilon/2)\Gamma^2(1+\epsilon/2)$. Omitting vanishing in the limit $\epsilon\to 0$ terms, we can perform integration over Feynman's variables $x_i$:
	\begin{align}
	\int_{0}^{1}dx_2\int_{0}^{1-x_2}dx_3 \frac{1}{(x_2(1-x_2))^{1+\epsilon/2}(1-x_3+a_j x_3)}&=-\frac{2\log (1+\gamma_j)}{\epsilon(1+\gamma_j)}-\frac{2\gamma_j}{1+\gamma_j}\frac{\partial}{\partial x} \ {}_3F_2\left (1,1,1;2,x;\frac{\gamma_j}{1+\gamma_j}\right ) \Bigl |_{x=1}\notag\\
	&+\frac{\partial}{\partial x} \ {}_3F_2\left (1,1,x;2,1;\frac{\gamma_j}{1+\gamma_j}\right ) \Bigl |_{x=1}+O(\epsilon) .
	\end{align}
Next, using the following property of a hypergeometric ${}_3 F_2$ function:
	\begin{equation}
\frac{\partial}{\partial x} \ {}_3F_2\left (1,1,1;2,x;\frac{\gamma}{1+\gamma}\right ) \Bigl |_{x=1} = -
\frac{\partial}{\partial x} \ {}_3F_2\left (1,1,x;2,1;\frac{\gamma}{1+\gamma}\right ) \Bigl |_{x=1} = 
\frac{(1+\gamma) \log^2 (1+\gamma)}{2\gamma} ,
	\end{equation}
we find
	\begin{equation}
	\widetilde{\mathcal{P}}_2^{\alpha_1\alpha_2;++;(2)} = - 8 \mu_2 \frac{t^2\,h^{2\epsilon}}{\epsilon^2} \sum_{j=0}^3 \Bigl [
	\ln (1+\gamma_j) -\frac{\epsilon}{4} \ln^2(1+\gamma_j)\Bigr ] .
	\label{eq2loopK2_P2++2}
	\end{equation}
	
The two-loop contribution to $\widetilde{\mathcal{P}}^{\alpha_1\alpha_2}_2(i\varepsilon_{n_1},i\varepsilon_{m_1})$ with $n_1 \geq 0$ and $m_1 < 0$ is equal to
	\begin{equation}
	\widetilde{\mathcal{P}}^{\alpha_1\alpha_2;(2)}_2(i\varepsilon_{n},i\varepsilon_{m}) = 
	-\frac{1}{4} \sum_{n_1m_1}\sum_{\beta_1\beta_2} \bigl\langle \bigl\langle 
	\spp[ w^{\alpha_1\beta_1}_{nm_1}\bar{w}^{\beta_1\alpha_1}_{m_1n}]\cdot
	\spp[ \bar{w}^{\alpha_2\beta_2}_{mn_1}{w}^{\beta_2\alpha_2}_{n_1m}] 
	\bigr\rangle \bigr\rangle  +\mu_2 \Bigl \langle \spp[ w^{\alpha_1\alpha_2}_{nm} \bar{w}^{\alpha_2\alpha_1}_{mn}]
	\Bigl [ S_{h,4}+\frac{1}{2} S_{h,3}^2\Bigr ] \Bigr \rangle  .
	\label{eqP2+-0}
	\end{equation} 
Here irreducible average is defined as usual, $\langle\langle A \cdot B \rangle \rangle \equiv \langle A B\rangle - \langle A\rangle \langle B \rangle$.  The expansion of the action $S_h$ to the third and forth orders in $W$ leads to the terms $S_{h,3}$ and $S_{h,4}$:
\begin{equation}
	S_{h,3}  =\sum_{j=0}^{3}\sum_{r=0;3}^{} \frac{\pi T \Gamma_j}{4} \sum_{\beta,n} \int d\bm{r} \Tr I^{\beta}_{n} t_{rj} W
	\Tr I^{\beta}_{-n}t_{rj}\Lambda W^2 ,	
	\end{equation} 
	and
	\begin{align}
	S_{h,4}  & = \frac{g}{128} \int_{q_1,\dots,q_4}\delta\left (\sum_{k=1}^4\bm{q_k}\right ) 
	\sum_{\beta_1,\dots,\beta_4}\sum_{n_{3,4},m_{3,4}}\spp [w^{\beta_1\beta_2}_{n_3m_3}(\bm{q_1}) \bar{w}^{\beta_2\beta_3}_{m_3n_4}(\bm{q_2})
	w^{\beta_3\beta_4}_{n_4m_4}(\bm{q_3}) \bar{w}^{\beta_4\beta_1}_{m_4n_3}(\bm{q_4})] \Bigl [ (\bm{q_1}+\bm{q_2})(\bm{q_3}+\bm{q_4}) 	\notag \\
	& +(\bm{q_1}+\bm{q_4})(\bm{q_2}+\bm{q_3})-2h^2 - \frac{16 Z_\omega}{g} (\Omega_{n_3m_3}^\varepsilon+\Omega_{n_4m_4}^\varepsilon)\Bigr ]  -\sum_{j=0}^{3}\sum_{r=0;3}^{}\frac{\pi T \Gamma_j}{16} \sum_{\beta,n} \int d\bm{r} \Tr I^{\beta}_{n}t_{rj} \Lambda W^2
	\Tr I^{\beta}_{-n}t_{rj}\Lambda W^2  .
	\end{align}
After the evaluation of averages in Eq.~\eqref{eqP2+-0} with the help of the Wick's theorem, we obtain
	\begin{gather}
	\widetilde{\mathcal{P}}^{\alpha_1\alpha_2;(2)}_2(i\varepsilon_{n},i\varepsilon_{m}) =  -2 \left [ \frac{16}{g} \int_q \mathcal{D}_q(i\Omega_{nm}^\varepsilon)\right ]^2 +2\mu_2\left (\frac{16}{g}\right )^2\int_{pq}\left [p^2+q^2+h^2+\frac{16Z_\omega}{g}\Omega_{nm}^\varepsilon\right ]\mathcal{D}_p(i\Omega_{nm}^\varepsilon)\mathcal{D}_q ^2(i\Omega_{nm}^\varepsilon)\notag \\
	+\mu_2\left (\frac{64}{g}\right )^2\sum_{j=0}^{3}\left (\sum_{\omega_n>\varepsilon_{n}}^{}+\sum_{\omega_n>-\varepsilon_{m}}^{}\right )\frac{\pi T \Gamma_j}{g}\int_{pq}\mathcal{D}_p^2(i\Omega_{nm}^\varepsilon)[p^2+h^2+q^2+\frac{16Z_\omega}{g}(\Omega_{nm}^\varepsilon+\omega_n)]\mathcal{D}_q(i\omega_n)\mathcal{D}_q^{(j)}(i\omega_n)\notag\\
	-\mu_2\left (\frac{64}{g}\right )^2\sum_{j=0}^{3}\sum_{\omega_n>0}\frac{2\pi T \Gamma_j}{g}\left [1-\frac{16\Gamma_j\omega_n}{g}\int_{pq}\mathcal{D}_{\bm{p}+\bm{q}}^{(j)}(i\omega_n)\right ]\mathcal{D}_q^2(i\Omega_{nm}^\varepsilon)\mathcal{D}_p(i\Omega_{nm}^\varepsilon+i\omega_n)\notag \\
	-\mu_2\left (\frac{64}{g}\right )^2\sum_{j=0}^{3}\left (\sum_{\epsilon_{n_1}>\omega_n>0}+\sum_{-\epsilon_{n_4}>\omega_n>0}\right )\frac{\pi T \Gamma_j}{g}\left [1-\frac{16\Gamma_j\omega_n}{g}\int_{pq}\mathcal{D}_{\bm{p}+\bm{q}}^{(j)}(i\omega_n)\right ]\mathcal{D}_q^2(i\Omega_{nm}^\varepsilon)\mathcal{D}_p(i\Omega_{nm}^\varepsilon-i\omega_n)
	\label{eq2loopK2_P2+-1}
	\end{gather}
	After the analytic continuation to the real energies in Eq. \eqref{eq2loopK2_P2+-1}, we obtain ($\Omega=E_1-E_2$)
	\begin{gather}
	\widetilde{\mathcal{P}}^{\alpha_1\alpha_2;+-;(2)}_2(E_1,E_2) =  -2 \left [ \frac{16}{g} \int_q \mathcal{D}^R_q(\Omega)\right ]^2 +2\mu_2\left (\frac{16}{g}\right )^2\int_{pq}\left [p^2+q^2+h^2-\frac{16Z_\omega i \Omega}{g}\right ]\mathcal{D}^R_p(\Omega)\mathcal{D}^{R2}_q(\Omega)\notag \\
	+\mu_2\left (\frac{64}{g}\right )^2\sum_{j=0}^{3}\frac{\Gamma_j}{4ig}\int_{pq} \int_{-\infty}^\infty d\omega \mathcal{D}^{R2}_p(\Omega)\left [p^2+h^2+q^2- \frac{16Z_\omega i}{g}(\Omega+\omega)\right ]\Bigl ( \mathcal{F}_{\omega-E_1}+ \mathcal{F}_{\omega-E_2}\Bigr )\mathcal{D}^{R}_q(\omega)\mathcal{D}^{R(j)}_q(\omega)\notag\\
	-\mu_2\left (\frac{64}{g}\right )^2\sum_{j=0}^{3} \frac{\Gamma_j}{2ig}\int_{pq}\int_{-\infty}^\infty d\omega \mathcal{B}_\omega \left [1+\frac{16\Gamma_ji\omega}{g}\mathcal{D}_{\bm{p}+\bm{q}}^{R(j)}(\omega)\right ]\mathcal{D}_q^{R2}(\Omega)\mathcal{D}^R_p(\Omega+\omega)\notag \\
	-\mu_2\left (\frac{64}{g}\right )^2\sum_{j=0}^{3} \frac{\Gamma_j}{4ig} \int_{pq} \int_{-\infty}^\infty d\omega \Bigl (2\mathcal{B}_\omega-\mathcal{F}_{\omega-E_1}-\mathcal{F}_{\omega-E_2}\Bigr )\left [1+\frac{16\Gamma_ji\omega}{g}\mathcal{D}_{\bm{p}+\bm{q}}^{R;(j)}(\omega)\right ]\mathcal{D}^{R2}_q(\Omega)\mathcal{D}^R_p(\Omega-\omega) .
	\label{eq2loopK2_P2+-1a}
	\end{gather}
	Here we introduce the bosonic distribution function $\mathcal{B}_\omega=\coth[\omega/(2T)]$. 
Next, we set energies and temperature to zero, $E_1=E_2=T=0$. Integrals can be evaluated in a standard way. We illustrate evaluation by the most complicated integral: 
	\begin{equation}
I = \int_{0}^{\infty}d\omega\,\omega \int_{pq}^{}\mathcal{D}_p^{R(j)}(\omega)\mathcal{D}_q^{R2}(0)\mathcal{D}_{\bm{p}+\bm{q}}^R(\omega)=-\frac{A_\epsilon\Gamma(1-\epsilon)h^{2\epsilon}}{\epsilon\Gamma^2(1-\epsilon/2)}\int_{0}^{1}\prod_{i=1}^{3}dx_i \delta\left (\sum_{i}^{}x_i-1\right )\frac{x_2(x_1 x_2+x_2 x_3+x_3 x_1)^{-1-\epsilon/2}}{((1+\gamma_j)x_1+x_3)^2} .
	\end{equation}
To integrate over Feynman's variable we use the following parametrization $x_3=\frac{1-u}{s+1}$, $x_2=\frac{s}{s+1}$, and $x_1=\frac{u}{s+1}$, with $0< u\leqslant 1$, and $0<s\leqslant \infty$. Then we obtain
	\begin{gather}
	I=\int_{0}^{1}du\ \frac{[u(1-u)]^{1-\epsilon/2}B(2,-1-\epsilon/2)}{(1+\gamma_j u)^2}\
	{}_2F_1(2,-\epsilon,1-\epsilon/2;1-u+u^2)  ,
	\end{gather}
where $B(a,b)$ is the Euler's beta-function. Using the asymptotic expansion of the hypergeometric function:
	\begin{equation}
	_2F_1(2,-\epsilon,1-\epsilon/2,1-u+u^2)\simeq 1-\epsilon\left (\frac{1-u+u^2}{u-u^2}-\ln(u-u^2)\right ) + O(\epsilon^2) ,
	\end{equation}
	we find
	\begin{equation}
	I\simeq -\frac{2}{\epsilon}\frac{2\gamma_j-(2+\gamma_j)\ln(1+\gamma_j)}{\gamma_j^3}- \frac{2}{1+\gamma_j}+\frac{2(2+\gamma_j)}{\gamma_j^3}\left (\liq(-\gamma_j)+2\ln(1+\gamma_j)+\frac{\ln^2 (1+\gamma_j)}{4} \right ) .
	\end{equation}
	As a total result, we obtain
	\begin{gather}
	\widetilde{\mathcal{P}}^{\alpha_1\alpha_2;+-;(2)}_2 = 32 \frac{t^2\,h^{2\epsilon}}{\epsilon^2} \Bigl ( -1+\mu_2 + \frac{\mu_2}{2} \epsilon \Bigr ) + 8\mu_2 \frac{t^2\,h^{2\epsilon}}{\epsilon^2}\sum_{j=0}^3\Bigl [
	2 f(\gamma_j )+ 3 \ln (1+\gamma_j) - \epsilon\frac{2+\gamma_j}{\gamma_j}  \Bigl ( \ln(1+\gamma_j) + \liq(-\gamma_j) \notag \\
	+ \frac{1}{4}\ln^2(1+\gamma_j)\Bigr ) \Bigr ] .
	\label{eq2loopK2_P2+-2}
	\end{gather}
Combining Eq.\eqref{eq2loopK2_P2++2} and \eqref{eq2loopK2_P2+-2}, we obtain the two-loop contribution to the irreducible correlation function 
	\begin{gather}
	\widetilde{\mathcal{K}}_2^{(2)}   = \frac{t^2\,h^{2\epsilon}}{4\epsilon^2} \Biggl \{ 4 - 2\mu_2(2+\epsilon) -2\mu_2 \sum_{j=0}^3 \Bigl [   f(\gamma_j) + 2 \ln(1+\gamma_j) \Bigr ] 
	-\mu_2 \epsilon \sum_{j=0}^3 \Bigl [ \ln(1+\gamma_j) + 2 f(\gamma_j) - c(\gamma_j) 
	\Bigr ] \Biggr \} .\label{eqK1_0}
	\end{gather}
	\end{widetext}	
	
The reducible correlation function can be obtained from the irreducible one with the help of the following relation: ${\mathcal{K}}_2^{(2)}=Z+\widetilde{\mathcal{K}}_2^{(2)}$. We introduce the quantity $m_2^\prime$ such that the relation ${\mathcal{K}}_2^{(2)} = Z m_2^\prime$ holds. Using Eqs. \eqref{eq:K2:1-app} and \eqref{eqK1_0}, we obtain Eq. \eqref{eq:m2l:ren} for $m_2^\prime$ with 
\begin{gather}
b_1 = \mu_2, \quad b_2 = 1-\mu_2 -\frac{\mu_2}{2} \sum_{j=0}^3 f(\gamma_j), \notag \\
 b_3 = \frac{\mu_2}{4}
\sum_{j=0}^3 c(\gamma_j)  .
\label{eq:bijk:app}
\end{gather}

In order to find the anomalous dimension of $m^\prime_2$, we will use the minimal subtraction scheme (see e.g. \cite{Amit-book}). We introduce dimensionless quantity $\bar{t} = t^\prime h^{\prime \epsilon}$ and, using Eqs. \eqref{eqS1} and \eqref{eq:m2l:ren}, express $t$ and $m_2$  as
\begin{gather}
t =  (h^\prime)^{-\epsilon} \bar{t} Z_t,\qquad 
 m_2= m_2^\prime 
Z_{m_2} .
\end{gather}
Although the interaction parameters $\gamma_j$ are renormalized at the one-loop level (see Refs. \cite{Finkelstein1990,BelitzKirkpatrick1994}), we do not need to take it into account since $b_1$ is independent of $\gamma_j$. To the lowest orders in $\bar{t}$ the renormalization parameters can be found as
\begin{equation}
Z_t= 1 + \frac{a_1}{\epsilon}\bar{t}, 
\label{eqZZZ2}
\end{equation}
and
\begin{equation}
Z_{m_2}^{-1} =  1 + \frac{b_1}{\epsilon}\bar{t} +
\frac{\bar{t}^2}{\epsilon^2} \Bigl [ b_2 + b_1 a_1 + \epsilon b_3 \Bigr ] .
\label{eqZZZ2}
\end{equation}
Now the anomalous dimension of $m_2^\prime$ can be derived in a standard manner from the condition that $m_2$ does not depend on the momentum scale $h^\prime$. In this way, one finds within two-loop approximation:
\begin{equation}
-\frac{d \ln m^\prime_2}{d y} = \zeta_2 =  b_1 t + 2 t^2 \left [ b_3 + \frac{1}{\epsilon} \left (b_2-\frac{b_1(b_1-a_1)}{2}\right ) \right ] .
\label{eqm2RG:app}
\end{equation}
The anomalous dimension \eqref{eqm2RG:app}
 has finite limit at $\epsilon \to 0$ provided the condition 
 \begin{equation}
 b_2 =b_1(b_1-a_1)/2
 \label{eq:cond:app}
 \end{equation}
  holds. This is equivalent to the algebraic equation for $\mu_2$ (see Eq. \eqref{eq:cons-cond}):
 \begin{equation}
 \mu_2^2+\mu_2-2=0.
 \end{equation}
The striking feature of this equation is its independence of the interaction parameters $\gamma_j$.  
There are two solutions $\mu_2=-2$ and $\mu_2=1$ which determine two bilinear in $Q$ pure scaling operators. With the help of Eq. \eqref{eq:bijk:app}, Eq. \eqref{eqm2RG:app} is reduced to Eq. \eqref{eqm2RG}.

\subsection{Operators with three $Q$s.}

The reducible operator $\mathcal{K}_3$ with three $Q$ matrices involves the following local operator 
	\begin{gather}
		\mathcal{P}_3^{\alpha_1\alpha_2\alpha_3}(i\varepsilon_{n},i\varepsilon_{m},i\varepsilon_k) = \langle  \spp Q_{nn}^{\alpha_1\alpha_1} \spp Q_{mm}^{\alpha_2\alpha_2} \spp Q_{kk}^{\alpha_3\alpha_3} \rangle  \notag\\
		 + \mu_{2,1}  \langle \spp \bigl [Q_{nm}^{\alpha_1\alpha_2} Q_{mn}^{\alpha_2\alpha_1} \bigr ]  \spp Q_{kk}^{\alpha_3\alpha_3}  \rangle \notag\\
		+\mu_3 \langle \spp \bigl [Q_{nm}^{\alpha_1\alpha_2} Q_{mk}^{\alpha_2\alpha_3}Q_{kn}^{\alpha_3\alpha_1} \bigr ] \rangle
		.
		\label{eqP3corr-tilde}
	\end{gather}
It is convenient to represent the reducible operator  $\mathcal{K}_3$ as a sum of the irreducible one, $\widetilde{\mathcal{K}}_3$, and the pure scaling operators with two $Q$ matrices (we denote them as
${\mathcal{K}}^{(-2)}_2$  and ${\mathcal{K}}^{(1)}_2$):
	\begin{equation}
		\mathcal{K}_3  = 
		\widetilde{\mathcal{K}}_3 + \sqrt{Z} \Bigl ( \mathcal{K}_2^{(-2)}+2 \mathcal{K}_2^{(1)}+
		\frac{\mu_2}{3} \bigl ( \mathcal{K}_2^{(1)}-\mathcal{K}_2^{(-2)}\bigr ) -2 Z \Bigr ) .
		\label{eq:K3:app1}
	\end{equation}
The irreducible correlation function $\widetilde{\mathcal{K}}_3$ has no one-loop contribution. The two-loop contribution is related to the operator $\mathcal{A}_3$. We find
	\begin{equation}
		\widetilde{\mathcal{K}}_3 = \frac{3\mu_3 t^2 h^{2\epsilon}}{2\epsilon^2}
\label{eq:irK3:app}
	\end{equation}
Using Eqs. \eqref{eq:irK3:app} and \eqref{eq:bijk:app}, from Eq. \eqref{eq:K3:app1} we obtain the relation $
\mathcal{K}_3 = Z^{3/2} m_3^\prime$ where $m_3^\prime$ is given by 
Eq. \eqref{eq:m2l:ren} with
\begin{gather}
b_1 = \mu_{2,1}, \quad b_2 = 3+\frac{3}{2} \mu_3-\mu_{2,1} -\frac{\mu_{2,1}}{2} \sum_{j=0}^3 f(\gamma_j), \notag \\
 b_3 = \frac{\mu_{2,1}}{4}
\sum_{j=0}^3 c(\gamma_j)  .
\label{eq:bijk1:app}
\end{gather}
Then the renormalizability condition \eqref{eq:cond:app} yields the following equation (see Eq. \eqref{eq:cons-cond}):
\begin{equation}
		 \mu_{2,1}^2+\mu_{2,1} = 3\mu_3 +6 .
		\label{eq:cond2:app}
\end{equation}
Again this equation is independent of the interaction parameters $\gamma_j$ and satisfied by the coefficients
$\mu_{2,1}$ and $\mu_3$ for noninteracting case from Table \ref{Tab}. 

\begin{widetext}
\subsection{Operators with four $Q$s.}
The reducible operator $\mathcal{K}_4$ with four $Q$ matrices involves the following local operator
\begin{gather}
		\mathcal{P}_4^{\alpha_1\alpha_2\alpha_3\alpha_4}(i\varepsilon_{n},i\varepsilon_{m},i\varepsilon_k,i\varepsilon_l)  = \langle  \spp Q_{nn}^{\alpha_1\alpha_1} \spp Q_{mm}^{\alpha_2\alpha_2}  \spp Q_{kk}^{\alpha_3\alpha_3}\spp Q_{ll}^{\alpha_4\alpha_4} \rangle  
		 + \mu_{2,1,1}  \langle \spp \bigl [Q_{nm}^{\alpha_1\alpha_2} Q_{mn}^{\alpha_2\alpha_1}\bigr ]  \spp Q_{kk}^{\alpha_3\alpha_3} \spp Q_{ll}^{\alpha_4\alpha_4} \rangle
		 \notag \\
		   + \mu_{2,2}  \langle \spp \bigl [Q_{nm}^{\alpha_1\alpha_2} Q_{mn}^{\alpha_2\alpha_1} \bigr ]  \spp \bigl [Q_{kl}^{\alpha_3\alpha_4} Q_{lk}^{\alpha_4\alpha_3}\bigr ]\rangle  
		+\mu_{3,1} \langle \spp \bigl [Q_{nm}^{\alpha_1\alpha_2} Q_{mk}^{\alpha_2\alpha_3} Q_{kn}^{\alpha_3\alpha_1}\bigr ] \spp Q_{ll}^{\alpha_4\alpha_4} \rangle		 + \mu_4  \langle \spp \bigl [Q_{nm}^{\alpha_1\alpha_2} Q_{mk}^{\alpha_2\alpha_3} Q_{kl}^{\alpha_3\alpha_4} Q_{ln}^{\alpha_4\alpha_1}\bigr ]  \rangle .
\label{eqP4corr-tilde} 
\end{gather}
The reducible operator  $\mathcal{K}_4$ can be written as a sum of the irreducible one, $\widetilde{\mathcal{K}}_4$, and the pure scaling operators with two and three $Q$ matrices (we denote the latter as
${\mathcal{K}}^{(-6)}_3$, ${\mathcal{K}}^{(-1)}_3$,  and ${\mathcal{K}}^{(3)}_3$):	
\begin{gather}
		\mathcal{K}_4  = \widetilde{\mathcal{K}}_4 + 3 Z^2 - 2 Z \left ({\mathcal{K}}_2^{(-2)} +2{\mathcal{K}}_2^{(1)}\right ) - \frac{\mu_{2,1,1}}{3} Z \left ({\mathcal{K}}_2^{(1)}- {\mathcal{K}}_2^{(-2)} \right )  
		+ \frac{4}{15}Z^{1/2}  
		\left ( {\mathcal{K}}_3^{(-6)}  + 9{\mathcal{K}}_3^{(-1)} +5{\mathcal{K}}_3^{(3)} \right ) 
		\notag \\ 
		+ \frac{\mu_{3,1}}{60} Z^{1/2} \left ( 4{\mathcal{K}}_3^{(-6)}  - 9{\mathcal{K}}_3^{(-1)}+5{\mathcal{K}}_3^{(3)} \right )
		+ \frac{\mu_{2,1,1}}{15} Z^{1/2}
		\left ( -2{\mathcal{K}}_3^{(-6)}- 3{\mathcal{K}}_3^{(-1)} +5{\mathcal{K}}_3^{(3)} \right )  .
		\label{eq:K4:app1}
	\end{gather}
\end{widetext}
The irreducible correlation function $\widetilde{\mathcal{K}}_4$ has no one-loop contribution. The two-loop contribution is related with the operator $\mathcal{A}_{2,2}$. We find
	\begin{equation}
		\widetilde{\mathcal{K}}_4 = \frac{\mu_{2,2} t^2 h^{2\epsilon}}{\epsilon^2}
		\label{eq:irK4:app}
	\end{equation}
Using Eqs. \eqref{eq:irK4:app} and \eqref{eq:bijk:app}, from Eq. \eqref{eq:K4:app1} we obtain the relation $
\mathcal{K}_4 = Z^{2} m_4^\prime$ where $m_4^\prime$ is given by 
Eq. \eqref{eq:m2l:ren} with
\begin{equation}
\begin{split}
b_1 &= \mu_{2,1,1}, \quad b_3 = \frac{\mu_{2,1,1}}{4}
\sum_{j=0}^3 c(\gamma_j)  ,  \\
b_2 &= 6+\mu_{2,2} + \frac{3}{2} \mu_{3,1}-\mu_{2,1,1} -\frac{\mu_{2,1,1}}{2} \sum_{j=0}^3 f(\gamma_j) .
\end{split}
\label{eq:bijk12:app}
\end{equation}
Then the renormalizability condition \eqref{eq:cond:app} yields the following equation (see Eq. \eqref{eq:cons-cond}):
\begin{equation}
		 \mu_{2,1}^2+\mu_{2,1} = 3\mu_{3,1} +2 \mu_{2,2}+12 \ .
		\label{eq:cond3:app}
\end{equation}
Again this equation is independent of the interaction parameters $\gamma_j$ and satisfied by the coefficients
$\mu_{2,1,1}$, $\mu_{3,1}$ and $\mu_{2,2}$ for noninteracting case from Table \ref{Tab}. 

\subsection{Operators with more than four $Q$s.}

The irreducible operator $\tilde{\mathcal{K}}_q$ with $q>4$ has no nonzero contributions within the two-loop approximation. Thus the reducible operator $\mathcal{K}_q$ can be fully expressed as a linear combination of pure scaling operators $\mathcal{K}^{(\mu)}_{q^\prime}$ with $q^\prime < q$. We use the following relations valid within two-loop approximation 
\begin{gather}
	\bigl \langle  \prod \limits_{j=1}^q\spp Q_{n_{k_j}n_{k_j}}^{\alpha_j\alpha_j}\bigr  \rangle \simeq\prod \limits_{j=1}^q\bigl \langle \spp Q_{n_{k_j}n_{k_j}}^{\alpha_j\alpha_j}\bigr  \rangle + \prod \limits_{j=1}^{q-2}\bigl \langle \spp Q_{n_{k_j}n_{k_j}}^{\alpha_j\alpha_j} \bigr \rangle \notag \\
	\times \frac{q(q-1)}{2}
	\bigl \langle \bigl \langle\spp Q_{n_{k_{q-1}}n_{k_{q-1}}}^{\alpha_{q-1}\alpha_{q-1}} \cdot\spp Q_{n_{k_q}n_{k_q}}^{\alpha_q\alpha_q} \bigr \rangle \bigr \rangle ,
	\label{eq:rel1:app}
\end{gather}
\begin{gather}
	\bigl \langle \spp \bigl [Q_{n_{k_1}n_{k_2}}^{\alpha_1\alpha_2} Q_{n_{k_2}n_{k_1}}^{\alpha_2\alpha_1} \bigr ]   \prod \limits_{j=3}^q\spp Q_{n_{k_j}n_{k_j}}^{\alpha_j\alpha_j} \bigr \rangle\simeq  \prod \limits_{j=1}^{q-2} \bigl \langle \spp Q_{n_{k_j}n_{k_j}}^{\alpha_j\alpha_j} \bigr  \rangle \notag \\
	\times
	\bigl \langle \spp \bigl [Q_{n_{k_{q-1}}n_{k_q}}^{\alpha_{q-1}\alpha_q} Q_{n_{k_q}n_{k_{q-1}}}^{\alpha_q\alpha_{q-1}} \bigr ]\bigr \rangle 
	+ (q-2) \prod \limits_{j=1}^{q-3}\bigl \langle \spp Q_{n_{k_j}n_{k_j}}^{\alpha_j\alpha_j} \bigr \rangle \notag \\
	\times \bigl\langle \bigl\langle \spp Q_{n_{k_{q-3}}n_{k_{q-3}}}^{\alpha_{q-3}\alpha_{q-3}}\cdot \spp \bigl [Q_{n_{k_{q-1}}n_{k_q}}^{\alpha_{q-1}\alpha_q} Q_{n_{k_q}n_{k_{q-1}}}^{\alpha_q\alpha_{q-1}} \bigr ]\bigr \rangle\bigr \rangle ,
\end{gather}
\begin{gather}
	\bigl \langle \spp \bigl [Q_{n_{k_1}n_{k_2}}^{\alpha_1\alpha_2} Q_{n_{k_2}n_{k_3}}^{\alpha_2\alpha_3}  Q_{n_{k_3}n_{k_1}}^{\alpha_3\alpha_1} \bigr ] 
	 \prod \limits_{j=4}^q\spp Q_{n_{k_j}n_{k_j}}^{\alpha_j\alpha_j} \bigr
	\rangle \simeq 
	 \prod \limits_{j=4}^q \notag \\
	\times \bigl \langle 
	\spp Q_{n_{k_j}n_{k_j}}^{\alpha_j\alpha_j} \bigr \rangle  \bigl \langle \spp \bigl [Q_{n_{k_1}n_{k_2}}^{\alpha_1\alpha_2} Q_{n_{k_2}n_{k_3}}^{\alpha_2\alpha_3}  Q_{n_{k_3}n_{k_1}}^{\alpha_3\alpha_1} \bigr ] \bigr \rangle ,
\end{gather}
\begin{gather}
	\bigl \langle \spp \bigl [Q_{n_{k_1}n_{k_2}}^{\alpha_1\alpha_2}Q_{n_{k_2}n_{k_3}}^{\alpha_2\alpha_3}  Q_{n_{k_3}n_{k_4}}^{\alpha_3\alpha_4}Q_{n_{k_4}n_{k_1}}^{\alpha_4\alpha_1} \bigr ] 
	 \prod \limits_{j=5}^q\spp Q_{n_{k_j}n_{k_j}}^{\alpha_j\alpha_j}
	\bigr \rangle \simeq \prod \limits_{j=5}^q\notag \\
	 \times \bigl \langle 
	\spp Q_{n_{k_j}n_{k_j}}^{\alpha_j\alpha_j} \bigr \rangle \bigl \langle \spp \bigl [Q_{n_{k_1}n_{k_2}}^{\alpha_1\alpha_2}Q_{n_{k_2}n_{k_3}}^{\alpha_2\alpha_3}  Q_{n_{k_3}n_{k_4}}^{\alpha_3\alpha_4}Q_{n_{k_4}n_{k_1}}^{\alpha_4\alpha_1} \bigr ] 
	\bigr \rangle ,
\end{gather}
and
\begin{gather}
	\bigl \langle \spp \bigl [Q_{n_{k_1}n_{k_2}}^{\alpha_1\alpha_2} Q_{n_{k_2}n_{k_1}}^{\alpha_2\alpha_1} \bigr ] \spp \bigl [ Q_{n_{k_3}n_{k_4}}^{\alpha_3\alpha_4}Q_{n_{k_4}n_{k_3}}^{\alpha_4\alpha_3} \bigr ] \prod \limits_{j=5}^q\spp Q_{n_{k_j}n_{k_j}}^{\alpha_j\alpha_j}
	\bigr
	\rangle \notag \\
	\simeq \prod \limits_{j=5}^q\bigl \langle 
	\spp Q_{n_{k_j}n_{k_j}}^{\alpha_j\alpha_j} \bigr \rangle \bigl \langle \spp \bigl [Q_{n_{k_1}n_{k_2}}^{\alpha_1\alpha_2} Q_{n_{k_2}n_{k_1}}^{\alpha_2\alpha_1} \bigr ] \notag \\
	\times \spp \bigl [ Q_{n_{k_3}n_{k_4}}^{\alpha_3\alpha_4}Q_{n_{k_4}n_{k_3}}^{\alpha_4\alpha_3} \bigr ] \bigr \rangle  .
	\label{eq:rel2:app}
\end{gather}
Using Eqs. \eqref{eq:rel1:app} - \eqref{eq:rel2:app}, we find 
\begin{widetext}
\begin{gather}
	\mathcal{K}_q =
	Z^{q/2} +\frac{q(q-1)}{6} Z^{(q-2)/2} \left ( {\mathcal{K}}_2^{(-2)}+ 2 {\mathcal{K}}_2^{(1)} - 3 Z \right ) + \frac{\mu_{2,1,\dots,1}}{3} Z^{(q-2)/2} \left ( {\mathcal{K}}_2^{(1)}- \tilde{\mathcal{K}}_2^{(-2)}  \right )
	+ \frac{(q-2)}{30}\mu_{2,1,\dots,1} Z^{(q-3)/2}\notag \\
	\times  \left (- 2{\mathcal{K}}_3^{(-6)}- 3 {\mathcal{K}}_3^{(-1)}+5 {\mathcal{K}}_3^{(3)}-  10 Z^{1/2}  {\mathcal{K}}_2^{(1)}+ 10 Z^{1/2}  {\mathcal{K}}_2^{(-2)}  \right )
	+ \frac{\mu_{3,1,\dots,1}}{60} Z^{(q-3)/2} \left (4{\mathcal{K}}_3^{(-6)}-9 {\mathcal{K}}_3^{(-1)}+5 {\mathcal{K}}_3^{(3)} \right ) \notag \\
	+ \mu_{4,1,\dots,1} Z^{(q-4)/2}  \left (-\frac{1}{105}{\mathcal{K}}_4^{(-12)}+\frac{2}{63} {\mathcal{K}}_4^{(-5)}+\frac{1}{180} {\mathcal{K}}_4^{(-2)} - \frac{2}{45}{\mathcal{K}}_4^{(1)} + \frac{1}{60} {\mathcal{K}}_4^{(6)}  \right ) \notag \\
	+ \mu_{2,2,1,\dots,1} Z^{(q-4)/2}  \left (\frac{1}{105}{\mathcal{K}}_4^{(-12)}-\frac{2}{63} {\mathcal{K}}_4^{(-5)}+\frac{7}{90} {\mathcal{K}}_4^{(-2)} - \frac{4}{45} {\mathcal{K}}_4^{(1)} + \frac{1}{30} {\mathcal{K}}_4^{(6)}  \right ) .
	\label{eq:Kq:app}
\end{gather}
\end{widetext}
From \eqref{eq:Kq:app} we obtain the relation $
\mathcal{K}_q = Z^{q/2} m_q^\prime$ where $m_q^\prime$ is given by 
Eq. \eqref{eq:m2l:ren} with coefficients $b_j$ defined in Eq. \eqref{eq:b123}. 
Then the renormalizability condition \eqref{eq:cond:app} is equivalent to Eq. \eqref{eq:cons-cond}.

\section{Background field method within one-loop approximation \label{App2}}

In this Appendix we provide details for a background field renormalization method of the operators $\mathcal{K}_q$ (with $q=3,4$) within one-loop approximation. As in the main text, we split the matrix field $Q$ into fast $\hat{Q}$ and slow $Q_0=\mathcal{T}_0^{-1} \Lambda \mathcal{T}_0$ components:  $Q = \mathcal{T}_0^{-1} \hat{Q} \mathcal{T}_0$. In this appendix we will treat the fast modes $\hat{Q}$ by expansion to the lowest nontrivial order in $W$. Using Eqs. \eqref{eq:prop:PH} and \eqref{eq:prop:PPT}, in the limit of $T\to 0$ we obtain the following well-known contraction rules:
\begin{gather}
\langle \Tr(AW)\Tr(BW)\rangle=Y \Tr \Bigl ( AB-\Lambda A\Lambda B-ACB^T C \notag\\
+\Lambda A \Lambda C B^T C\Bigr ),\quad Y=\frac{4}{g}\int_p \mathcal{D}_p(0)
\label{contr1}
\end{gather}
and
\begin{gather}
\langle \Tr AWBW \rangle=Y\Bigl (\Tr A  \Tr B -\Tr(\Lambda A)\Tr(\Lambda B)\notag \\
+\Tr ACB^TC -\Tr \Lambda A \Lambda C B^T C \Bigr ) .
\label{contr2}
\end{gather}
A few remarks are in order here. At first, Eqs. \eqref{contr1} and \eqref{contr2} are derived under assumption  that the fields $A$ and $B$ are slow whereas $W$ is fast. This allows us to neglect frequency dependence in diffusive modes. Secondly, due to the presence of interactions contractions \eqref{contr1} and \eqref{contr2} involve terms which are finite in the infrared and proportional to $T$. At $T=0$ such terms can be safely neglected and contractions  \eqref{contr1} and \eqref{contr2} are the same as in the noninteracting case.

\subsection{Operators with three $Q$s}

The basis for cubic in $Q$ operators involves the following three operators:
\begin{align}
\mathcal{A}_{1,1,1} &=\spp Q_{nn}^{\alpha_1\alpha_1} \spp Q_{mm}^{\alpha_2\alpha_2}  \spp Q_{kk}^{\alpha_3\alpha_3}, \notag \\
\mathcal{A}_{2,1} & = \spp [Q_{nm}^{\alpha_1\alpha_2} Q_{mn}^{\alpha_2\alpha_1 }] \spp Q_{kk}^{\alpha_3\alpha_3}, \\
\mathcal{A}_{3} & = \spp [Q_{nm}^{\alpha_1\alpha_2} Q_{mk}^{\alpha_2\alpha_3 } Q_{kn} ^{\alpha_3\alpha_1}] .
\end{align}
Within one-loop approximation \eqref{contr1} and \eqref{contr2} their transformation under the background field renormalization method is as follows:
\begin{gather}
\begin{pmatrix}
\mathcal{A}_{1,1,1}[Q]  \\
\mathcal{A}_{2,1}[Q] \\
\mathcal{A}_3[Q] 
\end{pmatrix} =\bigl [1+3(Z^{1/2}-1)\bigr ] \begin{pmatrix}
\mathcal{A}_{1,1,1}[Q_0]  \\
\mathcal{A}_{2,1}[Q_0] \\
\mathcal{A}_3[Q_0] 
\end{pmatrix} \notag \\
+ Y
\mathcal{M}_3
\begin{pmatrix}
\mathcal{A}_{1,1,1}[Q_0]  \\
\mathcal{A}_{1,2}[Q_0] \\
\mathcal{A}_3[Q_0] 
\end{pmatrix}  ,
\label{eq:bgfr:2}
\end{gather} 
where the matrix
\begin{equation}
\mathcal{M}_3=\begin{pmatrix}
0 & -6 & 0 \\
-1 & 1 & -4 \\
0 & -3 & 3 
\end{pmatrix} .
\end{equation}
The eigenvalues (with a minus sign) of the matrix $\mathcal{M}_3$ are $\lambda_2^{(1)} = -6$, $\lambda_2^{(2)} = -1$ and $\lambda_2^{(3)} = 3$. The corresponding eigenoperators are given as
\begin{align}
\mathcal{P}_{2}^{(-6)} &= \mathcal{A}_{1,1,1}-6\mathcal{A}_{2,1}+8 \mathcal{A}_3,  \notag \\
\mathcal{P}_{2}^{(-1)} &= \mathcal{A}_{1,1,1}-\mathcal{A}_{2,1}-2\mathcal{A}_3 ,  \\
\mathcal{P}_{2}^{(3)} &= \mathcal{A}_{1,1,1}+3\mathcal{A}_{2,1}+2\mathcal{A}_3, \notag .
\end{align} 
This implies exactly the same values of $\mu_{2,1}$ and $\mu_3$ as in Table \ref{Tab}.

\subsection{Operators with four $Q$'s}

For the operators with four $Q$'s we use the following basis:
\begin{align}
\mathcal{A}_{1,1,1,1}&=\spp Q_{nn}^{\alpha_1\alpha_1} \spp Q_{mm}^{\alpha_2\alpha_2}  \spp Q_{kk}^{\alpha_3\alpha_3} \spp Q_{ll}^{\alpha_4\alpha_4}, \notag \\
\mathcal{A}_{2,1,1}&= \spp \bigl[ Q_{nm}^{\alpha_1\alpha_2} Q_{mn}^{\alpha_2\alpha_1 } \bigr ] \spp Q_{kk}^{\alpha_3\alpha_3} \spp Q_{ll}^{\alpha_4\alpha_4}, \notag \\
\mathcal{A}_{3,1}&= \spp \bigl[ Q_{nm}^{\alpha_1\alpha_2} Q_{mk}^{\alpha_2\alpha_3 } Q_{kn}^{\alpha_3\alpha_1}\bigr] \spp Q_{ll}^{\alpha_4\alpha_4}, \notag \\
\mathcal{A}_{4}&=\spp \bigl [Q_{nm}^{\alpha_1\alpha_2} Q_{mk}^{\alpha_2\alpha_3}  Q_{kl}^{\alpha_3\alpha_4} Q_{ln}^{\alpha_4\alpha_1}\bigr] , \notag \\
\mathcal{A}_{2,2}&=\spp \bigl[ Q_{nm}^{\alpha_1\alpha_2} Q_{mn}^{\alpha_2\alpha_1 } \bigr ] \spp \bigl[ Q_{kl}^{\alpha_3\alpha_4} Q_{lk}^{\alpha_4\alpha_3 } \bigr ]  .
\end{align}
Within one-loop approximation \eqref{contr1} and \eqref{contr2} they transform under the background field renormalization method as follows:
\begin{gather}
\begin{pmatrix}
\mathcal{A}_{1,1,1,1}[Q]  \\
\mathcal{A}_{2,1,1}[Q] \\
\mathcal{A}_{2,2}[Q]\\
\mathcal{A}_{3,1}[Q]\\
\mathcal{A}_4[Q] 
\end{pmatrix} =\bigl [1+4(Z^{1/2}-1)\bigr ]\begin{pmatrix}
\mathcal{A}_{1,1,1,1}[Q_0]  \\
\mathcal{A}_{2,1,1}[Q_0] \\
\mathcal{A}_{2,2}[Q_0]\\
\mathcal{A}_{3,1}[Q_0]\\
\mathcal{A}_4[Q_0] 
\end{pmatrix} \notag \\
+ Y
\mathcal{M}_4 
\begin{pmatrix}
\mathcal{A}_{1,1,1,1}[Q_0]  \\
\mathcal{A}_{2,1,1}[Q_0] \\
\mathcal{A}_{2,2}[Q_0]\\
\mathcal{A}_{3,1}[Q_0]\\
\mathcal{A}_4[Q_0] 
\end{pmatrix} 
\label{eq:bgfr:4}
\end{gather} 
where the matrix
\begin{equation}
\mathcal{M}_4= \begin{pmatrix}
0 & -12& 0 & 0 & 0 \\
-1 & 1 & -2 & -8 & 0 \\
0 & -2& 2 & 0 & -8 \\
0 & -3 & 0 & 3 & -6 \\
0 & 0 & -2 & -4 & 6 
\end{pmatrix} .
\end{equation}
Eigenvalues (with a minus sign) of the matrix $\mathcal{M}_4$ are $\lambda_4^{(1)} = -12$, $\lambda_4^{(2)} = -5$, $\lambda_4^{(3)} = -2$, $\lambda_4^{(4)} = 1$, and $\lambda_4^{(5)} = 6$. The corresponding eigenoperators are
\begin{align}
\mathcal{P}_{4}^{(-12)} &= \mathcal{A}_{1,1,1,1}-12 \mathcal{A}_{2,1,1}+12\mathcal{A}_{2,2}+32 \mathcal{A}_{3,1}-48 \mathcal{A}_4,  \notag \\
\mathcal{P}_{4}^{(-5)} &= \mathcal{A}_{1,1,1,1}-5 \mathcal{A}_{2,1,1}+3\mathcal{A}_{2,2} +8 \mathcal{A}_{3,1}+6 \mathcal{A}_4,  \notag \\
\mathcal{P}_{4}^{(-2)} &= \mathcal{A}_{1,1,1,1}-2 \mathcal{A}_{2,1,1}+ 7\mathcal{A}_{2,2}-8 \mathcal{A}_{3,1}+2 \mathcal{A}_4,  \notag \\
\mathcal{P}_{4}^{(1)} &= \mathcal{A}_{1,1,1,1}+ \mathcal{A}_{2,1,1}-2\mathcal{A}_{2,2}- 2\mathcal{A}_{3,1}-4 \mathcal{A}_4,  \notag \\
\mathcal{P}_{4}^{(6)} &= \mathcal{A}_{1,1,1,1}+6 \mathcal{A}_{2,1,1}+3\mathcal{A}_{2,2}+8 \mathcal{A}_{3,1}+6 \mathcal{A}_4 .
\end{align} 
This implies exactly the same values of $\mu_{2,1,1}$, $\mu_{3,1}$ $\mu_{2,2}$ and $\mu_4$ as in Table \ref{Tab}.

\bibliography{paper-biblio}

\end{document}